\documentclass[reprint,amsmath,amssymb, aps]{revtex4-2}
\usepackage{amsmath,amssymb}
\usepackage{graphicx}
\usepackage{dcolumn}
\usepackage{bm}
\usepackage{hyperref}
\usepackage{xcolor}
\usepackage{caption}
\usepackage{subcaption}

\def\avg#1{{\langle #1\rangle }}
\def\kB{{k_{\rm B}}}
\def\xCo{{x_{\rm Co}}}
\def\xCu{{x_{\rm Cu}}}
\def\rmd{{\rm d}}

\begin{document}

\title{Simulated structure and thermodynamics of decagonal Al-Co-Cu quasicrystals}
\author{Yang Huang}
%\email{yangh2@andrew.cmu.edu}
\affiliation{
  Physics Department, Carnegie Mellon University.
}
\affiliation{
  University of Science and Technology of China, Hefei 230026, China
}
\affiliation{
  Suzhou Institute for Advanced Research, University of Science and Technology of China, Suzhou 215213, China
}
\author{Michael Widom}
%\email{widom@cmu.edu}
\affiliation{
  Physics Department, Carnegie Mellon University.
}
\author{M. Mihalkovi\v c }
%\email{fyzimiha@savba.sk}
\affiliation{
  Inst. of Physics, Slovak Academy of Sciences, 84511 Bratislava, Slovakia
}

\date{\today}

\begin{abstract}
  Atomic structures of Al-Co-Cu decagonal quasicrystals (QCs) are investigated using empirical oscillating pair potentials (EOPP) in molecular dynamic (MD) simulations that we enhance by Monte Carlo (MC) swapping of chemical species and replica exchange. Predicted structures exhibit planar decagonal tiling patterns and are periodic along the perpendicular direction. We then recalculate the energies of promising structures using first-principles density functional theory (DFT), along with energies of competing phases. We find that our $\tau$-inflated sequence of QC approximants are energetically unstable at low temperature by at least 3 meV/atom. Extending our study to finite temperatures by calculating harmonic vibrational entropy, as well as anharmonic contributions that include chemical species swaps and tile flips, our results suggest that the quasicrystal phase is entropically stabilized at temperatures in the range 600-800K and above. It decomposes into ordinary (though complex) crystal phases at low temperatures, including a partially disordered B2-type phase. We discuss the influence of density and composition on QC phase stability; we compare the structural differences between Co-rich and Cu-rich quasicrystals; and we analyze the role of entropy in stabilizing the quasicrystal, concluding with a discussion of the possible existence of ``high entropy'' quasicrystals.
  
\end{abstract}
\maketitle

\section{Introduction}
Since the icosahedral quasicrystal Al$_{86}$Mn$_{14}$ was reported~\cite{Shechtman1984}, many new quasicrystals have been found, including other icosahedral phases~\cite{Tsai2013} plus layered phases~\cite{Steurer2004} with planar quasiperiodicity and perpendicular periodic stacking~\cite{Kuo88,Tsai89}. Decagonal quasicrystals exhibit 10-fold rotational symmetry in-plane, with stacking periodicities that are multiples of approximately 4~\AA. The symmetry space group has been controversial with reports of P10$_5$/mmc~\cite{Steurer1990}, $P\bar{10}m2$~\cite{Taniguchi2008} and $P5m1$~\cite{Yang2022}.  At compositions such as Al$_{65}$Co$_{17.5}$Cu$_{17.5}$ experiments find the quasicrystal to be thermodynamically stable over the temperature range 973-1350K~\cite{Dong_1991}. At lower temperatures it relaxes to a microcrystalline structure consisting of periodic approximants of the quasicrystal.

Structure modeling of $d-$AlCoCu has been aided by real-space imaging with high resolution transmission electron microscopy (HRTEM)~\cite{Hiraga1991}, including high-angle annular dark-field (HAADF) methods~\cite{Taniguchi2008,Yang2022} that highlight atoms of large atomic number~\cite{Pennycook2002}. Aperiodic tiling patterns and atomic occupation of quasicrystals are revealed directly in real space~\cite{Steurer2004}, supplementing the spatially averaged information provided by diffraction refinement.

Theoretical structure models have been proposed that are based directly on the experimental data~\cite{Steurer1990,Burkov1991,Burkov1993} or that include considerations of interatomic interactions~\cite{Mihalkovic2002,Cockayne1998}. Decagonal clusters feature prominently many models. Overlaps of these clusters~\cite{Gummelt1996} create shapes such as hexagons, boats and stars, leading in general to hexagon-boat-star-decagon (HBSD) tilings, and these tilings emerge at multiple length scales~\cite{Mihalkovic2002,Gu2006}. With growing computer power we can now perform full ab-initio total energy calculations for systems of several hundred atoms or more. In addition, improved methods for fitting interatomic interaction potentials~\cite{Mihalkovic2012} and advanced simulation techniques~\cite{Swendsen1986} allow calculation of entropic contributions to the free energy of quasicrystals and their competing phases~\cite{Mihalkovic2020}.

In the following, we first present our computational methods, then we discuss the global Al-Co-Cu phase diagram with a focus on the Al-rich region that contains the decagonal quasicrystal and also many competing crystal phases. Next, we present simulated structures, starting from small approximants that yield a structure that lies only 3 meV/atom above the convex hull of competing phases, and continuing to successivly larger approximants that reach up to 1475 atoms. We combine first principles total energy calculations with harmonic and anharmonic contributions to the free energy to show that the quasicrystal phase is stabilized by entropy at temperatures above approximately 600-800K. Atomic vibrations and phason fluctuations combine to achieve thermodynamic stability. Finally, we examine the possible existence of ``high entropy'' quasicrystals.

\section{Methods and Models}
\subsection{Quasicrystal approximants}
Quasicrystals lack translational symmetry, which makes it difficult to calculate properties of real infinite quasicrystals. To overcome this difficulty, we approach the infinite quasicrystal structure via periodic crystalline approximants (QCAs~\cite{eh85}) with large unit cells. The unit cells of our approximants are monoclinic, consisting of a rhombic prism with interior angles $2\pi/5$ and $3\pi/5$ at the base, allowing us to reproduce $5-$ and $10-$fold symmetry in the $xy$ plane with stacking periodicity along the $z$ direction.  Although we seek to model a phase with $\approx$4~\AA~ periodicity, we allow for local $\approx$8~\AA~ structural motifs, so we take $c=8.16$~\AA, and occasionally $c=16.32$~\AA. In-plane lattice constants of rhombus-based approximants relate to $a_0$ via~\cite{Steurer2004}
\begin{equation}
  \label{eq:taun}
  a_n=\frac{2\sqrt{3-\tau}}{5}\tau^{n+1}a_0
\end{equation}
where $\tau=(\sqrt{5}+1)/2=1.618\dots$ is the Golden Mean and we take $a_0=3.792$~\AA. Our approximant sizes range from $a_3=12.223$~\AA~ up to $a_6=51.777$~\AA.

\subsection{Hybrid Monte Carlo/molecular dynamics with replica exchange}
To efficiently optimize quasicrystal structures, replica exchange simulations are performed at fixed density and composition on multiple replicas~\cite{Swendsen1986,Mihalkovic2020} selected in a geometric sequence of 24 temperatures ranging from $T=100K$ to $T=1205K$. Between replica swaps we carry out 50 cycles of Monte Carlo chemical species swaps iterating over all atoms, and 1000 molecular dynamic steps (4.07fs per step) to ensure that each replica thoroughly samples its equilibrium ensemble at each temperature.

\begin{figure}[tbhp]
  \centering
  \includegraphics[height=.3\textwidth]{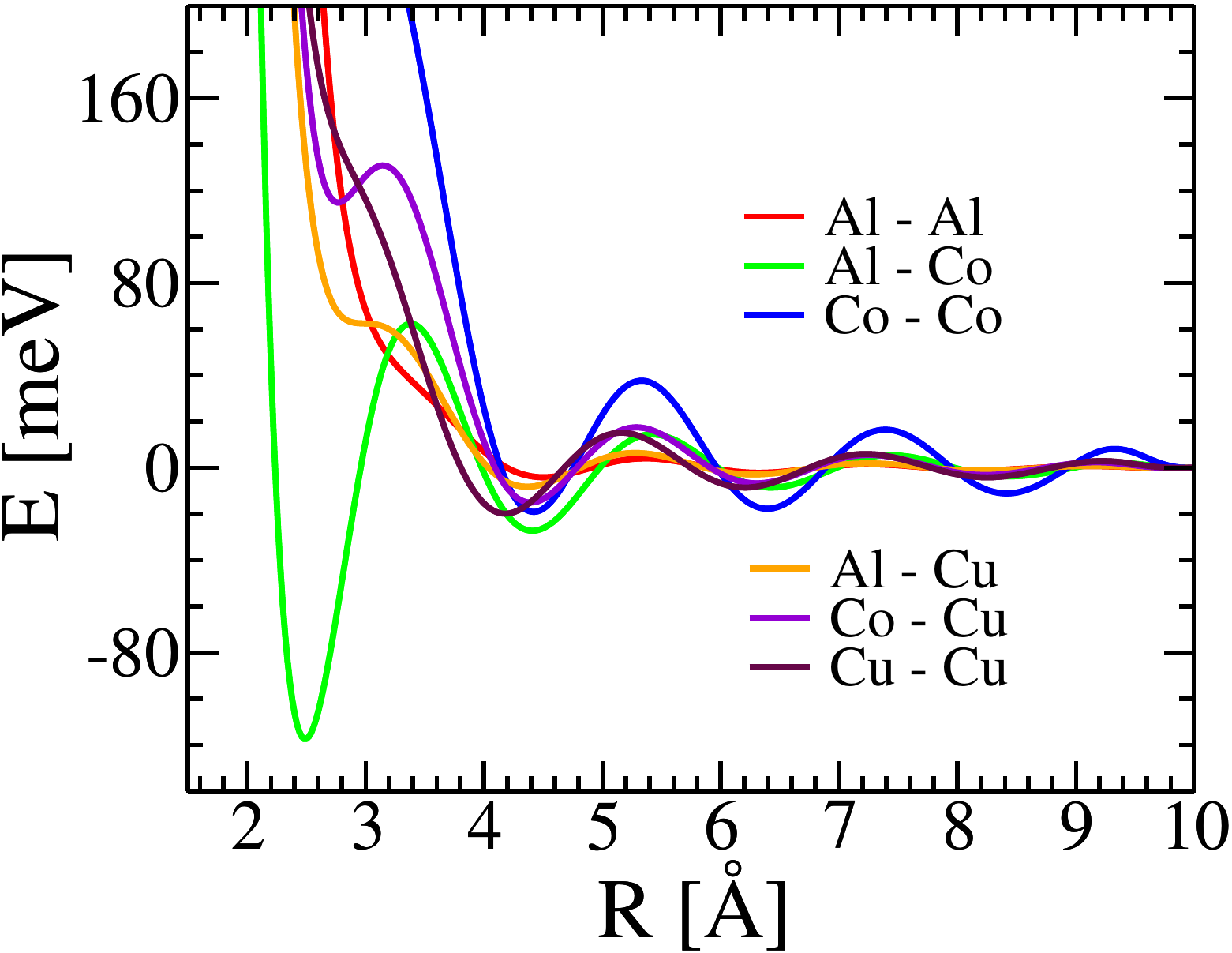}
  \caption{Empirical oscillating pair potentials for Al-Co-Cu.}
  \label{fig:EOPP}
\end{figure}

The simulations use empirical oscillating pair potentials (EOPP~\cite{Mihalkovic2012})
\begin{equation}
  \label{eq:eopp}
  V(r)=\frac{C_1}{r^{\eta_1}}+\frac{C_2}{r^{\eta_2}}\cos\left(kr+\phi\right),
\end{equation}
where the parameters are fitted to energies and forces of Al-Co-Cu alloys calculated from first principles. Values of the parameters and parity plots of fitted energies and forces are given in Appendix~\ref{app:EOPP}. Fig.~\ref{fig:EOPP} shows the six different pair potentials. Note the strong first minimum in Al-Co that lies close to the near-neighbor bond length of the AlCo.cP2 (Strukturbericht B2), and also the first minimum in Co-Co near 4.5~\AA~ that we will show is another important length scale. When $R$ is large, the potentials exhibit Friedel-like oscillations~\cite{Friedel1958} at frequencies that in principle~\cite{HafnerHam} depend on the Fermi wavenumber $k_F$. Because the Fermi wavenumber is a function of the electron density, application of this pair potential is intended to be limited to compositions and atomic volumes of the structures employed in the fit.

\subsection{First-principles calculations}
Although pair potentials are fast and sufficiently accurate for Monte Carlo and molecular dynamics simulations at elevated temperatures, they are not sufficiently accurate to predict the precise low temperature stable structures. For this purpose we perform first-principle calculations with electronic density functional theory as implemented in the Vienna Ab initio Simulation Package (VASP)~\cite{Kresse96}. We use augmented plane wave potentials~\cite{PAW} in the Perdew-Burke-Ernzerhof (PBE~\cite{PBE}) generalized gradient exchange correlation functional. All structures are fully relaxed (lattice parameters and coordinates) with a high electronic $k$-point density until their energies are converged to within $1$meV/atom. After relaxation we carry out a static calculation using tetrahedron integration to obtain the most accurate ground state energy.

\subsection{Vibrational and electronic free energies}

Vibrational free energy is calculated in the quantum harmonic approximation using Phonopy~\cite{phonopy}. Force constants are computed within VASP for crystal phases and for our smallest (83-atom) quasicrystal approximant, but we employ EOPP-derived forces for larger approximants. The accuracy of the vibrational density of states within EOPP is illustrated in Appendix~\ref{app:EOPP}. We neglect thermal expansion, as it is onerous to calculate from first principles for large unit cell low symmetry structures such as quasicrystal approximants; additionally our EOPP potentials are designed for use at fixed density. Electronic free energy is calculated based on the T=0K electronic density of states with a self-consistently determined temperature-dependent chemical potential. See Ref.~\cite{,Mihalkovic2020,WidomJMR2018} for more details of both the vibrational and electronic free energy methods.

\subsection{Anharmonic free energy}

The  free energy has additional contributions from phonon anharmonicity, chemical species swapping, and phason fluctuations, that we collectively label as anharmonic free energy~\cite{Mihalkovic2020}. Because these cannot be accurately calculated within density functional theory, we estimate the anharmonic contribution based on EOPP simulations, which is fast but less accurate.

To calculate anharmonic free energies, hybrid MC/MD with replica exchange is used to compute average energy $U$ (without kinetic contribution) and heat capacity $C_v = (\avg{U^2}-\avg{U}^2)/\kB T^2$.  Because the simulation is purely classical, and we already have the quantum harmonic free energy from Phonopy, we subtract the harmonic contributions for average energy and heat capacity to obtain the anharmonic contributions,
\begin{equation}
  \label{eq:Ua}
  U^a=U-\frac{3}{2}N k_{\rm B}T; \quad C_v^a=C_v-\frac{3}{2}N k_{\rm B},
\end{equation}
and the anharmonic relative entropy
\begin{equation}
  \label{eq:Sa}
  S^a(T)-S^a(T_0)=\int_{T_0}^T\,d\tau \frac{C_v}{\tau}
\end{equation}
assuming $S^a(T_0)=0$ for $T_0=100K$, and finally we obtain the anharmonic free energy
\begin{equation}
  \label{eq:Fa}
  G^a=U^a-TS^a
\end{equation}
This method is applied to the quasicrystal phase and to its primary crystalline competitors.

\subsection{HAADF simulations}

High aperture angular dark field (HAADF) electron microscopy images are simulated to compare our simulated structures with experiment. We compute these by simulating QCA structures sampled every 1ps over a total of 100ps. Atoms of these approximants are weighted by the squares of their atomic numbers and then projected into pixels of size 0.2~\AA~ in the simulated HAADF image, followed by a Gaussian smearing of 0.5~\AA. Pixel values are then displayed with an inverted gray scale in order to mimic the experimental images.

 \section{Calculation Results}
 
\subsection{Convex hull}
\label{sec:QC0K}

\begin{figure}[htpb]
  \centering
  \includegraphics[width=.4\textwidth]{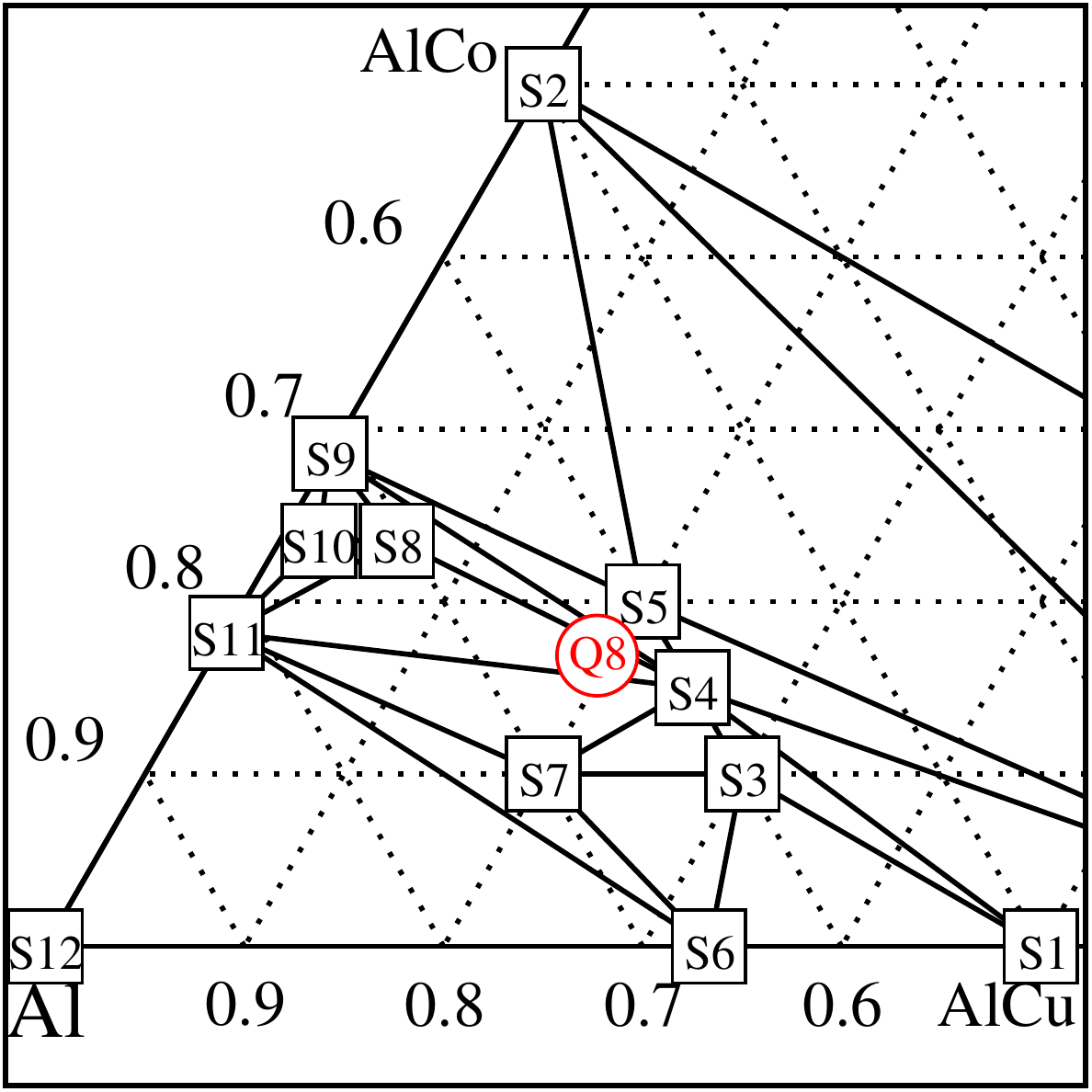}
  \caption{  \label{fig:Hull}
    Convex hull of Al-Co-Cu showing our optimized quasicrystal approximant and its competing phases. Structure information is summarized in Tab.~\ref{tab:Hull}.The horizontal axis is Al-Cu with Al-Co on the slant. A closeup with additional quasicrystal enthalpies is given in appendix~\ref{app:Hull-closeup}.}
\end{figure}

Formation enthalpies per atom \cite{Mihalkovi2004} of ${\rm Al}_x{\rm Co}_y{\rm Cu}_z$ are calculated based first principles T=0K energies per atom relative to the composition-weighted average of the constituent elements~\cite{Mihalkovi2004} via
\begin{equation}
  \label{eq:dH}
  \Delta H=E({\rm Al}_x{\rm Co}_y{\rm Cu}_z)-x E({\rm Al})-y E({\rm Co})-z E({\rm Cu}).
\end{equation}
The convex hull of the enthalpies predicts the phase diagram in the limit of low temperature, with vertices of the hull identifying the predicted stable phases (see Fig.~\ref{fig:Hull}). The stability of a phase at 0K is indicated by its relative energy $\Delta E$ with respect to the convex hull. We were unable to find any large quasicrystal approximant on the convex hull, however we were able to find a small 83-atom approximant with a low $\Delta E=3$ meV/atom.

\begin{table}[htpb]
  \centering
  \begin{ruledtabular}
    \begin{tabular}{llllll}
      index&Composition&Pearson symbol&$\Delta H$ [meV/atom]\\
      \hline
      S1&AlCu&mC20&-218.02\\
      S2&AlCo&cP2&-600.85\\
      S3&Al$_{12}$Co$_{2}$Cu$_{6}$&hP5&-309.45\\
      S4&Al$_{12}$Co$_{3}$Cu$_{5}$&hP5&-363.00\\
      S5&Al$_{12}$Co$_{4}$Cu$_{4}$&hP5&-404.11\\
      S6&Al$_2$Cu&cF12&-180.99\\
      S7&Al$_7$CoCu$_2$&tP40&-280.16\\
      S8&Al$_{72}$Co$_{24}$Cu$_6$&mC102&-420.81\\
      S9&Al$_5$Co$_2$&hP28&-461.49\\
      S10&Al$_{76}$Co$_{24}$Cu$_2$&mC102&-407.06\\
      S11&Al$_9$Co$_2$&mP22&-331.31\\
      S12&Al&cF4&0\\
      Q8&Al$_{53}$Co$_{14}$Cu$_{16}$&DQC& $\Delta E$=3.0
    \end{tabular}
  \end{ruledtabular}
  \caption{\label{tab:Hull} Formation enthalpies of structures on the convex hull, and our lowest $\Delta E$ QCA.}
\end{table}

The stable binary Al-Co phases are Al$_9$Co$_2$.mP22 (S11), Al$_5$Co$_2$.hP28 (S9), and AlCo.cP2 (S2). Not shown are Al$_{13}$Co$_4$ phases with Pearson types mC102 and oP102. These are well-known approximants of the Al-Co-based decagonal quasicrystal phases, but they lie slightly above the convex hull according to their DFT total energies~\cite{Mihalkovic2007}. However, they are stabilized by small additions of Cu  as shown in structures (S10) and (S8). Their structures are described below in Sect.~\ref{sec:mC102}. The AlCo.cP2 (B2, $\beta$) phase is closely related to the decagonal quasicrystal and may be considered as its periodic average structure~\cite{Steurer2000,Steurer2004}.

The Al$_{12}$M$_8$ phases (S3, S4, and S5) are based on 2x2x2 supercells of the $\tau_3$ phase with Pearson type hP5. They belong to a sequence of so-called $\tau$-phases~\cite{vanSande1978,VANTENDELOO1989,widom2000} that are vacancy-ordered variants of cP2, and are further discussed in Appendix~\ref{app:tau}. Inspection of Fig.~\ref{fig:Hull} shows that the $\tau$-phases will be the most important competitors to the quasicrystal at low temperatures, though the high configuraitonal entropy of B2 makes it an important competitor at high temperatures.

\subsection{Small approximants}

\subsubsection{(Al,Cu)$_{13}$Co$_4$.mC102}
\label{sec:mC102}

% {\bf (xx: is it Al$_{76}$Co$_{24}$Cu$_2$? Al$_{72}$Co$_{24}$Cu$_5$ is not shown in the table. Also note that the actual composition of Al$_{76}$Co$_{24}$Cu$_2$ is Al$_{76}$Co$_{24}$Cu$_{1.9}$ with $0.1$ vacancy. We make it up to $2$ in order to have a smaller cell.)} FIXED

\begin{figure}[htpb]
  \centering
    \includegraphics[width=.28\textwidth]{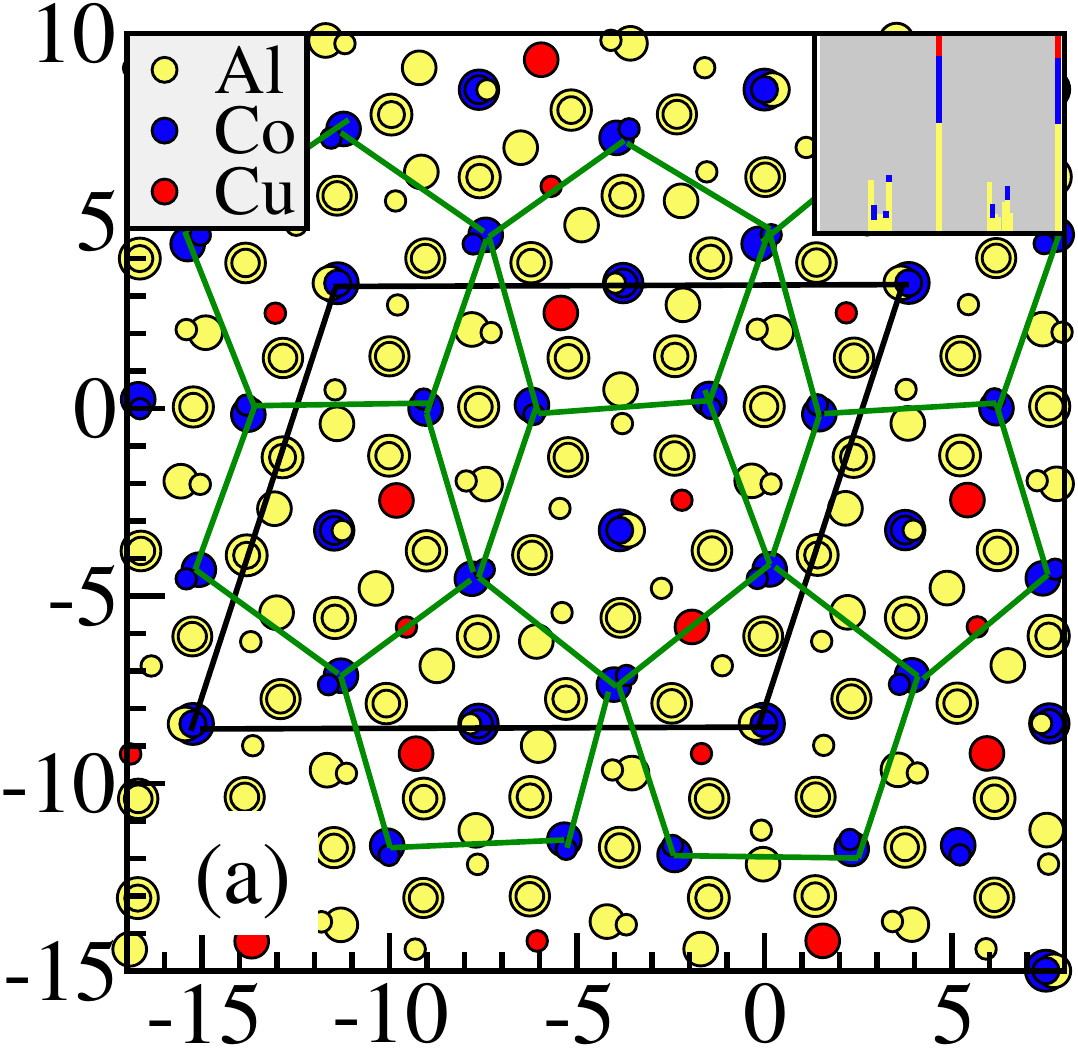}
    \includegraphics[width=.18\textwidth]{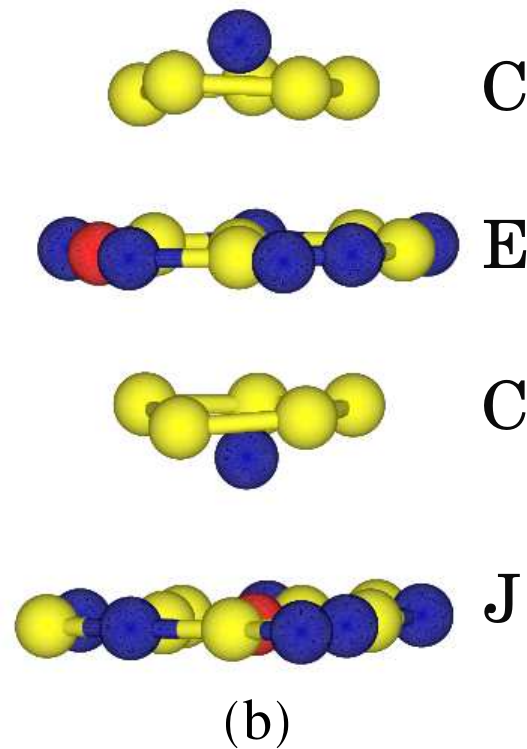}
  \caption{
    \label{fig:mC102}
    (a) VASP-relaxed structure of Al$_{72}$Co$_{24}$Cu$_6$.mC102. Atom sizes increase with greater depth. Inset shows depth profile. (b) A typical pentagonal bipyramid (PB) showing (top-to-bottom) top cap (C), equator (E), bottom cap (C), and junction layer (J). Cu substitutions for Co occur in the flat layers.}
\end{figure}

Extensions of Al$_{13}$Co$_4$.mC102 into the Al-Co-Cu ternary are reported in Ref.\cite{grushko_freiburg_1992}. Relaxed structures of Al$_{72}$Co$_{24}$Cu$_6$ (S8) and Al$_{76}$Co$_{24}$Cu$_2$ (S10) lie on the convex hull as shown in Fig.~\ref{fig:Hull}. Fig.~\ref{fig:mC102} presents the structure of Al$_{72}$Co$_{24}$Cu$_6$.mC102 in which we substituted Cu atoms for Al atoms at positions of Wyckoff class {\it 2a}. It has an 8~\AA~ periodicity with alternating flat and puckered layers as shown in the inset. The structure is built around pentagonal bipyramids (PBs~\cite{Henley-PBP}). The flat ``equatorial'' layer consists of a large pentagon of Co atoms alternating with a pentagon of Al atoms, and centered by a single Al atom. The flat ``junction'' layer differs from the equatorial layer due to its interior decoration with multiple Al atoms placed in an asymmetric and context-dependent manner. Cu atoms may substitute for Al in the flat layers. The puckered layers consist of mirror image PB ``caps'', which are Al$_5$ pentagons with a vertically offset centering Co atom. The mC102 structure periodically arranges four PBs, two of which are shifted vertically by 4~\AA~ so that each flat layer contains two equators and two junctions, yielding an average 4~\AA~ periodicity.

%The PB consists of a large Co pentagon alternating with Al atoms and centered by Al. This flat equatorial layer is capped above and below by mirror image Al pentagons and Co atoms. The entire unit is joined with its repeats along the $c$-axis by flat junction layers.

%The equator and junction layers both are defined by a large Co pentagon alternating with Al (or Cu) atoms. The equator is centered by a single Al atom, while the junction layers hold triangles of Al (or Cu) atoms. A periodic arrangement of $4$ pentagons is shown in the $xy$-plane. Two of these are PB equators, and the other two are PB junction layers, because the equators are shifted vertically by $4$\AA~ in height from the other two.

\subsubsection{Ternary approximants}
In contrast to the (Al,Cu)$_{13}$Co$_4$.mC102 approximant, which extends a binary phase into the ternary composition space, $d$-AlCoCu is a genuine ternary phase that spans a broad range of compositions~\cite{Grushko1993_LowCu,Grushko1993_HighCu,widom2000} from approximately Al$_{70}$Co$_{20}$Cu$_{10}$ (Co-rich) to Al$_{60}$Co$_{10}$Cu$_{20}$ (Cu-rich). We optimize the structures by performing MC/MD simulations with EOPP potentials then relax a selection of low temperature structures with VASP. This is repeated for varying numbers of atoms and composition in an effort to find the lowest possible energy. We find the most energetically favorable structures occur at TM-rich compositions with nearly equal Co and Cu concentrations, as presented in Appendix~\ref{app:Hull-closeup}.

\begin{figure*}[htpb]
  \centering
  \includegraphics[width=.32\textwidth]{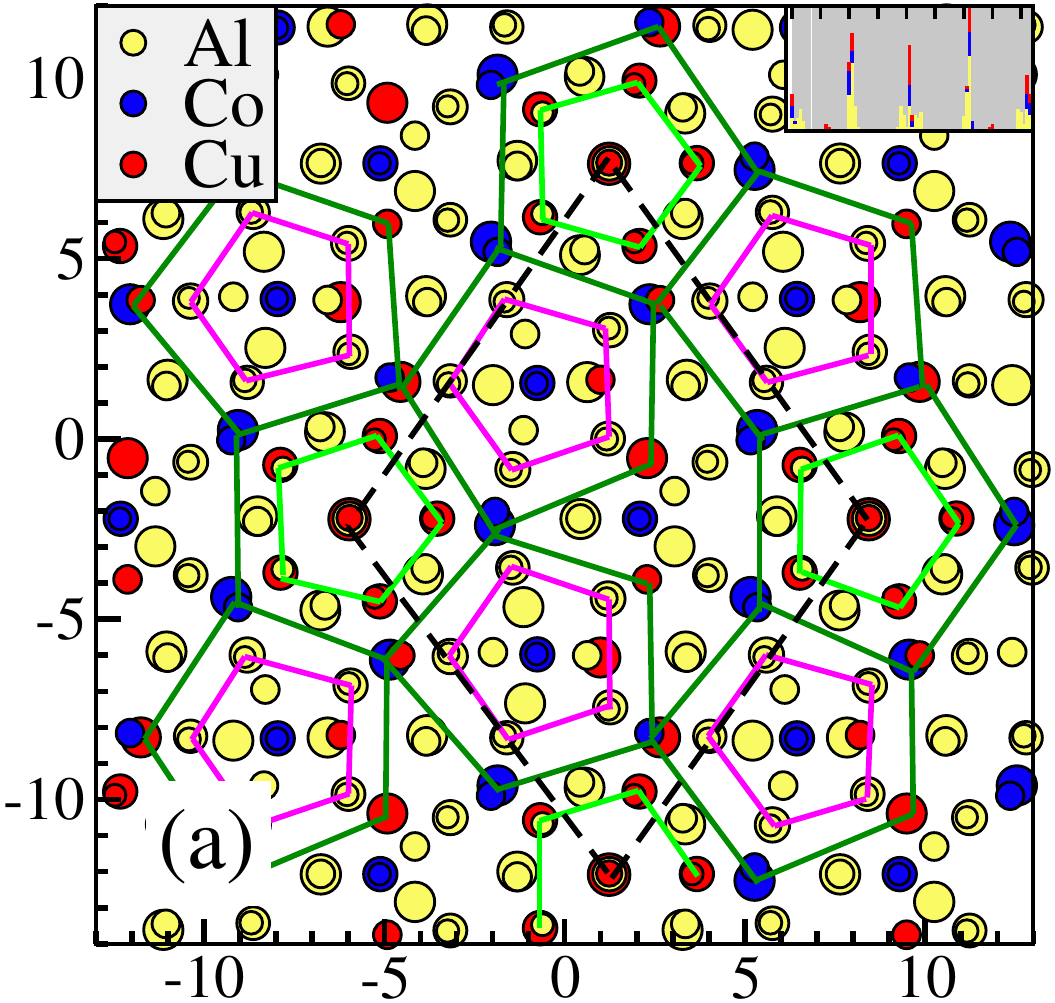}
  \includegraphics[width=.32\textwidth]{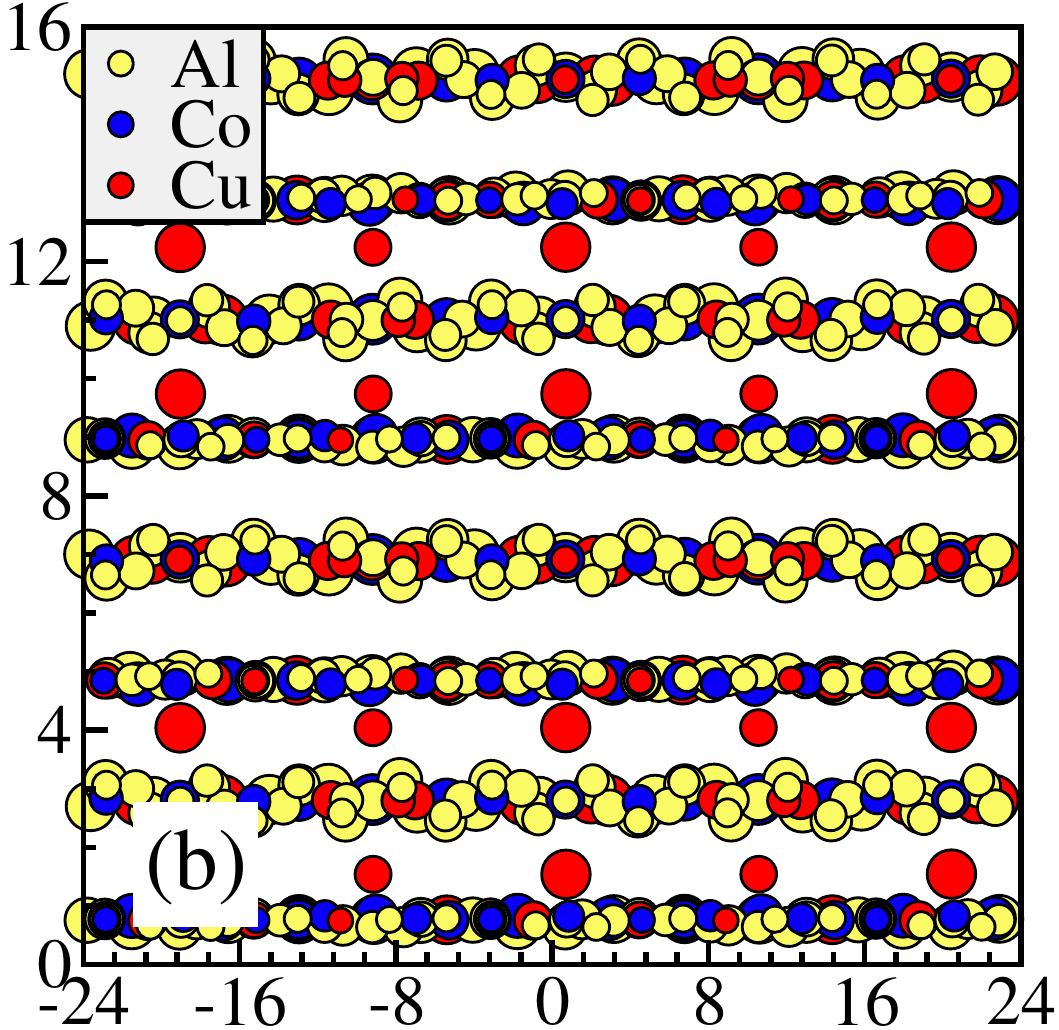}
  \includegraphics[width=.20\textwidth]{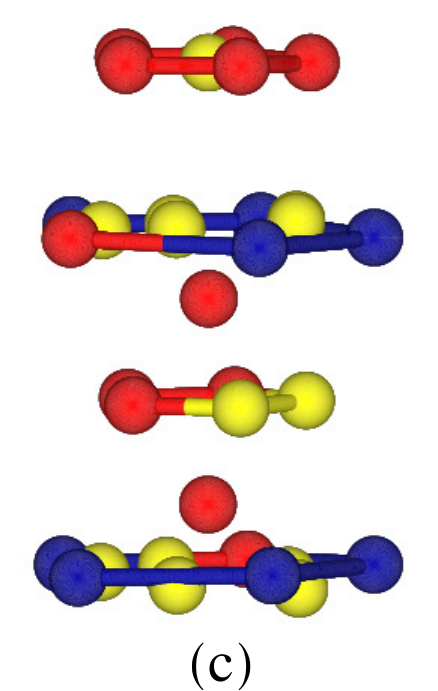}
  \caption{\label{fig:83} (a) 83-atom QC approximant Al$_{53}$Co$_{14}$Cu$_{16}$ (Q8) with edge length $a_3=12.223$~\AA~ and $c=8.2$~\AA. Atom sizes increase with greater depth; inset shows depth profile. Leftward-pointing (pink) pentagons highlight Al$_5$Co PB caps. Rightward-pointing (light green) pentagons highlight Cu-rich PB-like motifs. (b) side view revealing mid-layer Cu atoms. (c) Cu-rich PB-like motif with a dumbell of mid-layer Cu atoms.
}
\end{figure*}

Our energetically best structure, Al$_{53}$Co$_{14}$Cu$_{16}$ (Q8) contains 83 atoms with rhombus edge length $a_3=12.223$~\AA, and a slight excess of Cu relative to Co. It lies at $\Delta E=3.0$ meV/atom above the convex hull and has TM concentrations $\xCo=0.169$ and $\xCu=0.193$.
% with relaxed density $\rho=0.0711$~at./\AA$^3$.
As shown in Fig.~\ref{fig:83}, it exhibits 8~\AA~ periodicity and alternates puckered layers with nearly flat layers. In addition to ordinary PBs similar to those of mC102, new Cu-rich PB-like motifs appear with modified caps. The Al pentagons become Cu-rich, and the Co capping atoms are replaced alternately with a single Al, and with a mid-layer Cu dumbell. Thus the bottom cap becomes a miniature PB, with an Al-Cu pentagon equator of edge length 3.2~\AA~ sandwiched between the pair of dumbell Cu atoms.

The PB and PB-like motifs are highlighted in pink and green, respectively, in Fig.~\ref{fig:83}a. These two types of PBs are also found in the next larger approximant with 217 atoms and rhombus edge length $a_4=19.777$~\AA~  (not shown). It has composition Al$_{138}$Co$_{39}$Cu$_{41}$ ($\xCo=0.179$, $\xCu=0.188$) and lies at $\Delta E = 8.8$ meV/atom above the convex hull.
% Five pink pentagons arrange to form a larger pentagon that lies on the inner ring of a structure known as a Burkov cluster~\cite{Burkov1991}. The entire unit cell can be represented as a ``lightbulb'' tiling~\cite{Lehyani2003}.
In both the $a_3$ and $a_4$ approximants, the pentagons locating standard Co-rich  PBs point opposite to the pentagons locating Cu-rich PB-like clusters. Alternation of pentagon orientation is forced by the geometry of edge-sharing pentagons. Thus we conclude that it is energetically favorable to alternate Co-rich and Cu-rich PB motifs.

% \begin{figure}
%   \includegraphics[width=0.45\textwidth]{qc_218-atom_238.pdf}
%   \caption{\label{fig:218} 218-atom approximant.  Atom sizes increase with greater depth; inset shows depth profile. }
% \end{figure}

\subsection{Larger approximants}

We next studied larger $a_5$ and $a_6$ structures containing approximately 569 and 1490 atoms, respectively. Our practice of seeking low temperature structures from replica exchange MC/MD simulations proved problematic because our simulations at targeted compositions and densities ended up transforming into combinations of B2-like structures (with large vacancy concentrations, see App.~\ref{app:tau}) with Co-rich regions of quasicrystal-like order at the B2 grain boundaries. Similar microstructures have been observed experimentally~\cite{He1988,Zhang1992}. 

% \begin{figure}
%   \includegraphics[width=0.45\textwidth]{CoPhase_569-atom_100K.pdf}
%   \caption{\label{fig:coex}
%     Coexistence of B2-like structures and Co-rich regions of
%     quasicrystal-like order at T=100K from a 569-atom approximant simulation. NEED TO DECIDE IF WE WISH TO SHOW THIS FIGURE.
%   }
% \end{figure}

We found that complete 569-atom $a_5$ quasicrystal structures briefly reappear throughout the runs, and we selected several of these for first principles relaxation (see Table~\ref{tab:Hull-closeup}).
% Our lowest energy structure at composition Al$_{360}$Co$_{106}$Cu$_{103}$ (TM concentrations $\xCo=0.186$ and $\xCu=0.181$ with relaxed density $\rho=0.0719$ at./\AA$^3$) lay at $\Delta E = 11.4$ meV/atom above the convex hull and is illustrated in Fig.~\ref{fig:569}.
When we apply our computational methods to $a_6$ QCAs with 1490 atoms, no QC phase is found at Al$_{922}$Co$_{282}$Cu$_{286}$, which is similar in density and composition to our low-energy $a_5$ 569-atom QCA. Instead, the system irreversibly decomposes into B2-like structures.

\begin{figure}[htpb]
  \centering
  % \begin{subfigure}[b]{0.24\textwidth}
  %   \centering
    \includegraphics[width=.49\textwidth]{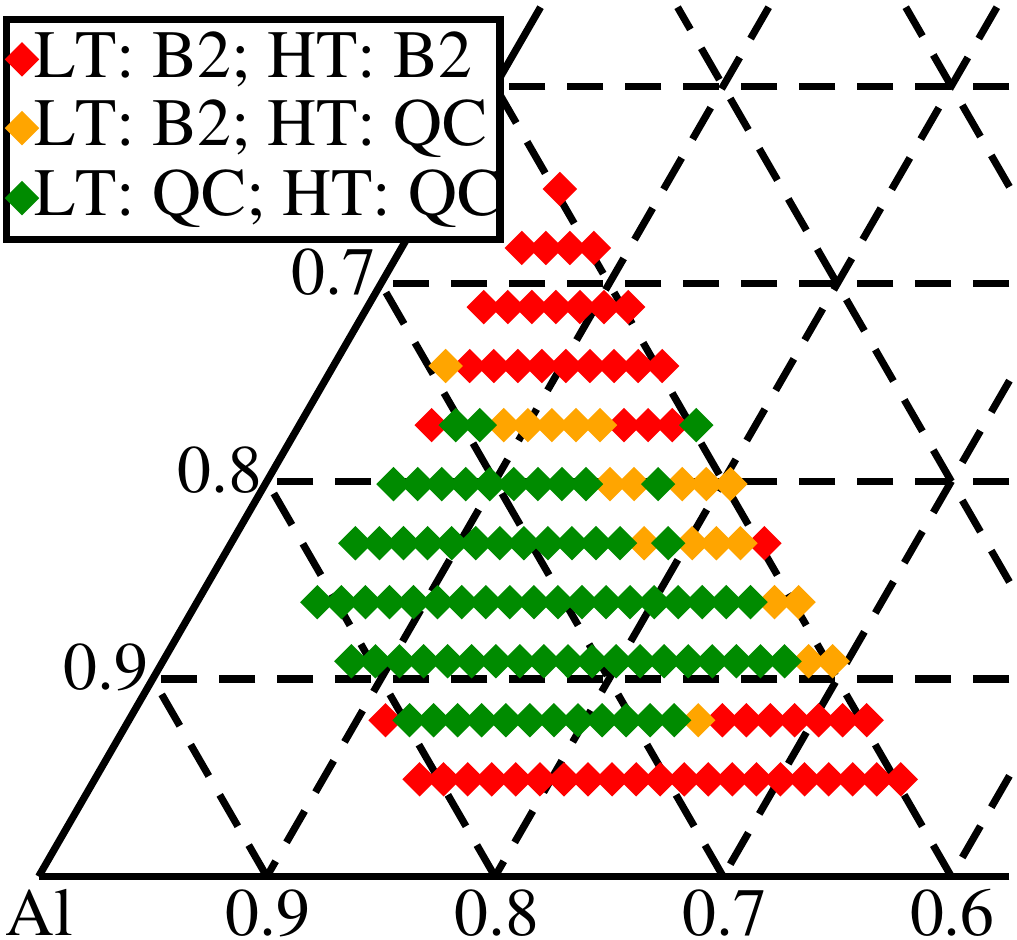}
  %   \caption{\label{fig:1475-atom_a}
  %     569-atom QCAs}
  % \end{subfigure}
  % \begin{subfigure}[b]{0.24\textwidth}
  %   \centering
  %   \includegraphics[width=\textwidth]{QC_1475-atom_fast-search_mod.pdf}
  %   \caption{\label{fig:1475-atom_b}
  %     1475-atom QCAs}
  % \end{subfigure}
  \caption{  \label{fig:fast-search}
    Final structures of 569-atom $a_5$ QCA after replica exchange simulations over wide composition ranges. Each diamond represents a simulation at its composition. Dark green compositions end with QC at both low and high temperatures. Orange diamonds are B2 at low temperatures and QC at high temperatures. Red diamonds are B2 at all temperatures.}
  \end{figure}

Smaller QCAs are more strongly favored by the constraints of the approximant cell size and shape than larger approximants are~\cite{CellConstrained}. To increase the chance of obtaining a large QCA, we explore compositions that are more Al-rich in order to be further from the region of B2 stability. To find the optimal QC-forming compositions (as opposed to the most energetically favorable compositions), we perform simulations that cover a wide composition range for $\tau^4\times 83=569$-atom $a_5$ QCAs at density similar to our optimal 83-atom $a_3$. Each simulation is initialized with a QCA at its target composition. Simulations are halted if they irreversibly transform to B2, and otherwise are continued for long runs up to 10ps of MD and $2.5\times 10^6$ MC swap attempts/atom. We divide the final results into three categories -- B2 at all temperatures, B2 at low temperature and QC at high temperature, and QC at all temperatures.

Final results for $a_5$-inflated QCAs with 569 atoms are summarized in Fig.~\ref{fig:fast-search}. We observe a wide Al-rich region of stable QC phase surrounded by TM-rich B2-favoring compositions. The QC phase extends further in the low-Al direction on the Cu-rich side~\cite{Grushko1993_LowCu,Grushko1993_HighCu,widom2000}.  No QCAs were stable for 1490-atom $a_6$ systems at our usual density. Hence we introduced vacancies and ran new trial searches with 1475 atoms. The increased entropy created by vacancies favors the QC over B2, and the QC phases persisted at low and high temperature over a wide range of compositions, especially those that were Al-rich. However, the lower density generally {\em increases} the DFT total energy per atom, so we do not expect these structures to lie near the convex hull.

  \subsection{TM-composition dependence}
  
\subsubsection{Structure of Cu-rich QCAs}
\begin{figure*}[htpb]
  \includegraphics[width=.55\textwidth]{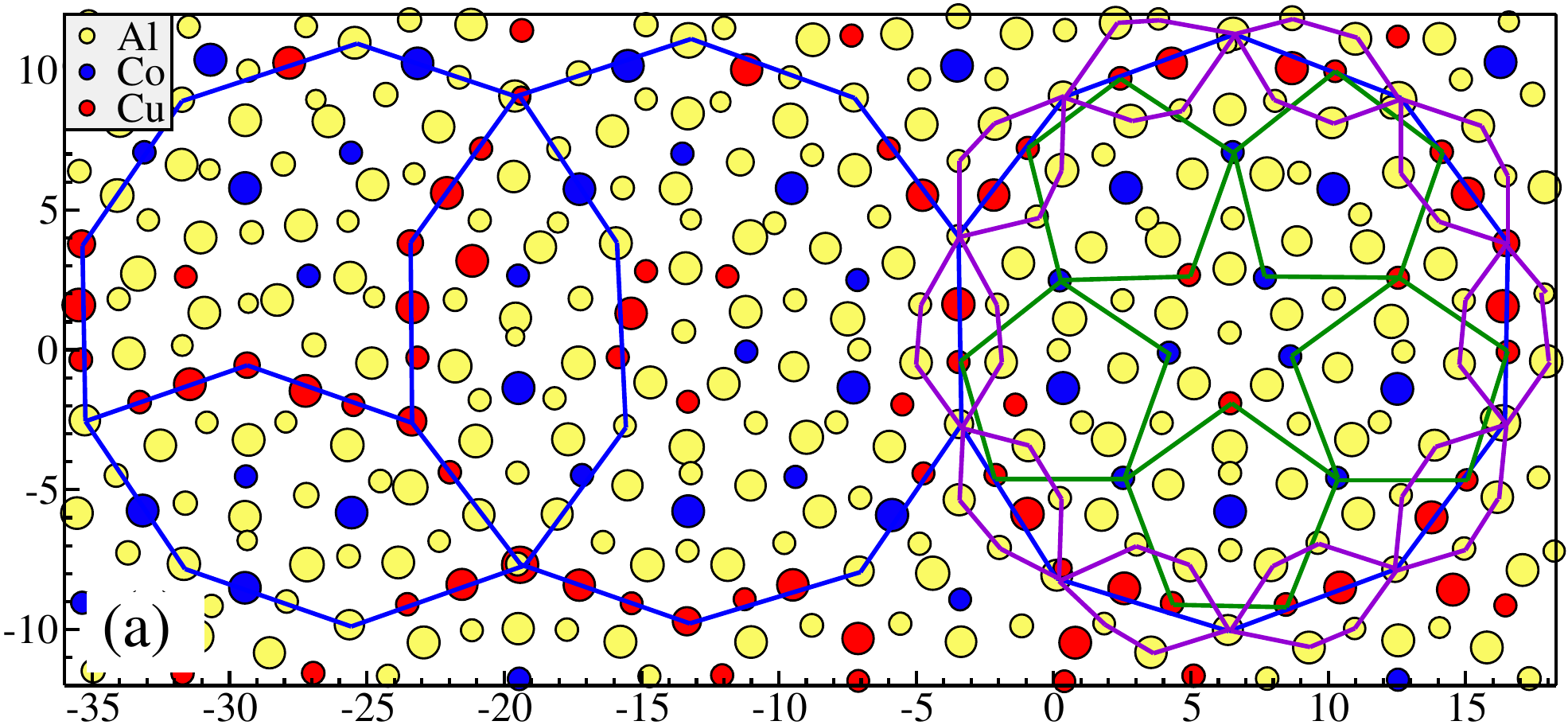}
  \includegraphics[width=.35\textwidth]{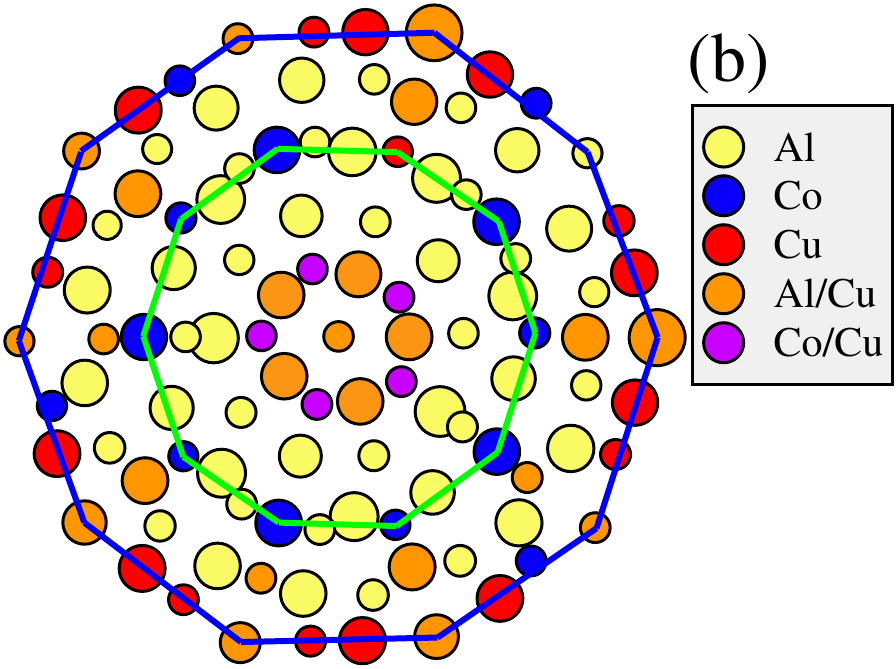}
  \caption{(a) A 4~\AA~ slab of our 1475-atom Cu-rich structure showing a 20~\AA~ Burkov cluster (blue), PB caps (dark green), and overlapping 20~\AA~ cluster motifs (blue).  Atom sizes increase with greater depth.
    (b) Simulated chemical occupancy model in the Burkov cluster (blue) and its inner 13~\AA~ cluster (light green). Dark green pentagons identify PB motifs. Purple hexagons highlight Cu pairs on Burkov cluster edges.}
  \label{fig:Cu-rich-atomic}
\end{figure*}

Burkov\cite{Burkov1991,Burkov1993} described the Cu-rich QC in terms of two-layered decagonal clusters with $\approx$20~\AA~ diameters and space group P10$_5$/mmc with centrosymmetry. These two layers are rotated by 36$^{\circ}$ leading to 4~\AA~ stacking periodicity, as shown in Fig.~\ref{fig:Cu-rich-atomic}b. The innermost ring of the Burkov cluster is a TM decagon with atoms alternating in height. The second shell is a 13~\AA~ decagon with TM atoms on vertices and Al atom pairs on edges. The outermost shell is a decagon with Al atoms on vertices and TM atom pairs on edges. A covering of the plane is achieved by allowing overlapping clusters.

However experimental HAADF results find that the 20~\AA~ cluster has a five-fold symmetry $P5m1$ without centrosymmetry\cite{Taniguchi2008,Yang2022}. It is also suggested that the 20~\AA~  cluster has 8~\AA~ periodicity along its symmetry axis because of interior PBs as shown in Fig.~\ref{fig:Cu-rich-atomic}(a). All evidences imply a different tiling pattern which are reflected in two aspects:  1) Adjacent clusters may have two opposite orientations which explains the global ten-fold symmetry ; 2) The intersection of two 20\AA~ clusters breaks the five-fold symmetry of the innermost shell and decomposes the decagon into a boat and a pair of hexagons~\cite{Cockayne1998}.

\begin{figure*}[htpb]
    \centering
    \begin{tabular}{cc}
        \begin{tabular}{c}
            \begin{subfigure}{0.69\textwidth}
              \centering
              \smallskip
              \smallskip
              \smallskip
              \includegraphics[width=\textwidth]{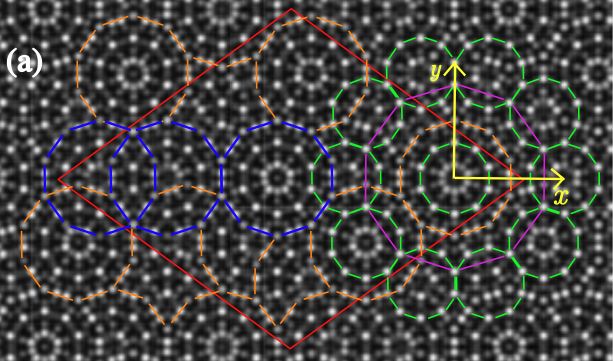}
              \label{fig:11c}
            \end{subfigure} 
        \end{tabular}
        &
          \begin{tabular}{c}
            \begin{subfigure}[t]{0.2\textwidth}
                \centering
                \includegraphics[width=\textwidth]{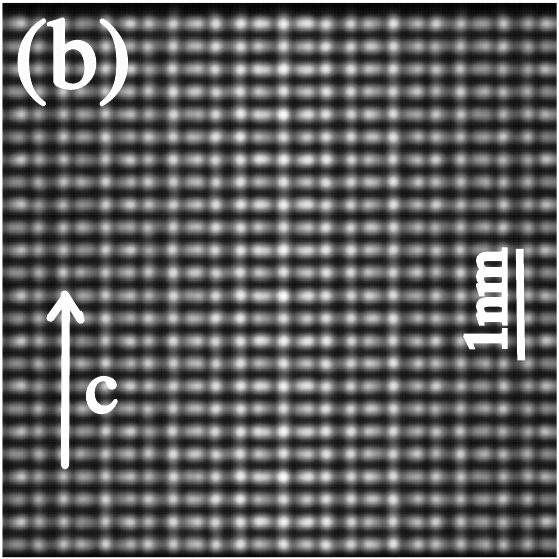}\\
            \end{subfigure} 
            \\
            \begin{subfigure}[t]{0.2\textwidth}
                \centering
                \includegraphics[width=\textwidth]{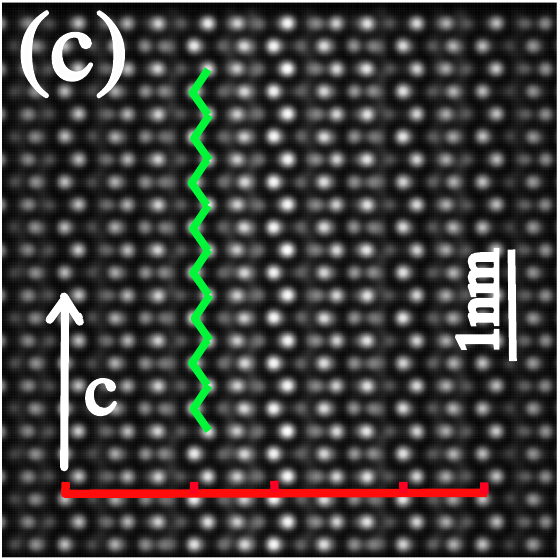}
            \end{subfigure}
        \end{tabular}
    \end{tabular}
  \caption{  \label{fig:Cu-rich-HAADF}
    (a) A simulated HAADF image of a 1475-atom Cu-rich $a_3$ QCA with composition Al$_{1040}$Co$_{161}$Cu$_{274}$. 8~\AA~ projection averaged over 100ps at T=809K. Orange hexagon-boat-star-decagon (HBSD) tiling has edge length 6.5~\AA. Green 13~\AA~ decagons assemble into a large superdecagonal ring. Red rhombus represents our QCA ($a=51.77$\AA). (b) A simulated HAADF image projecting in the $x$ direction (50\AA$\times$50\AA). (c) A simulated HAADF image when projected in the $y$ direction (50\AA$\times$50\AA).}
\end{figure*}

Fig.~\ref{fig:Cu-rich-atomic}a displays a typical 20~\AA~ cluster from one of our Cu-rich Al$_{940}$Co$_{205}$Cu$_{330}$ structures at low temperature. The innermost shell is an Al-centered TM-atom pentagon with Al atoms on edges. These five-fold 20~\AA~ clusters allow two opposite ({\em i.e.} 36$^\circ$ rotated) orientations defined by the internal TM pentagons. Our cluster model is supported by experimental HAADF images~\cite{Taniguchi2008,Yang2022} where at the center of a cluster is a bright pentagon whereas the Burkov cluster model predicts a bright decagon. General rules of site occupancy are summarized in Fig.~\ref{fig:Cu-rich-atomic}b. The Al sites of the innermost shell mix with Cu, and the Co sites also mix with Cu. However, the distinction between Al-containing and Co-containing pentagons is preserved. Few substitutions occur in the second (Co-rich, 13~\AA) shell. Vertices of the outermost shell become mixed Al/Cu, while the edges are mostly Cu pairs.

\begin{figure}[htpb]
  \centering
  \includegraphics[width=.45\textwidth]{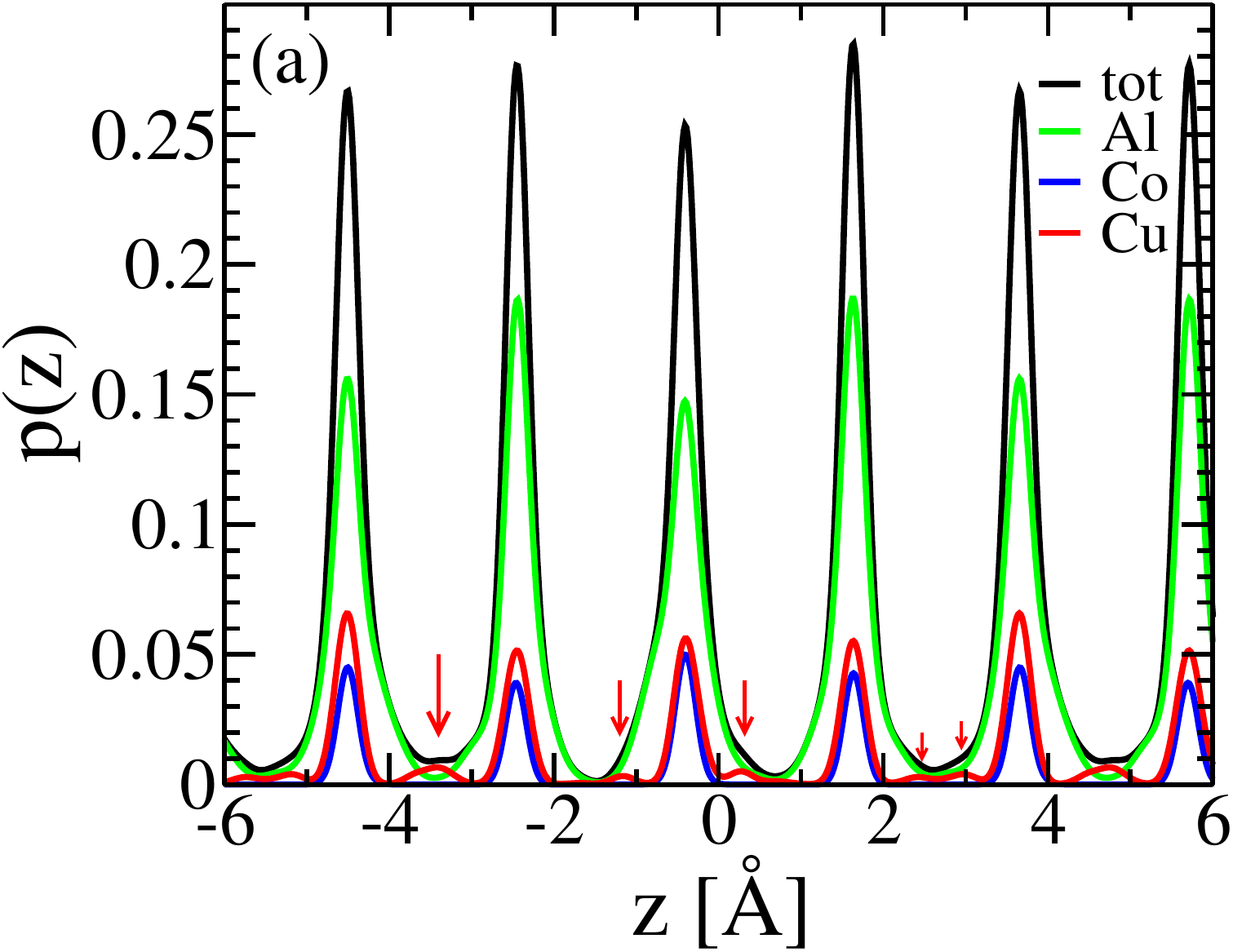}
  \includegraphics[width=.45\textwidth]{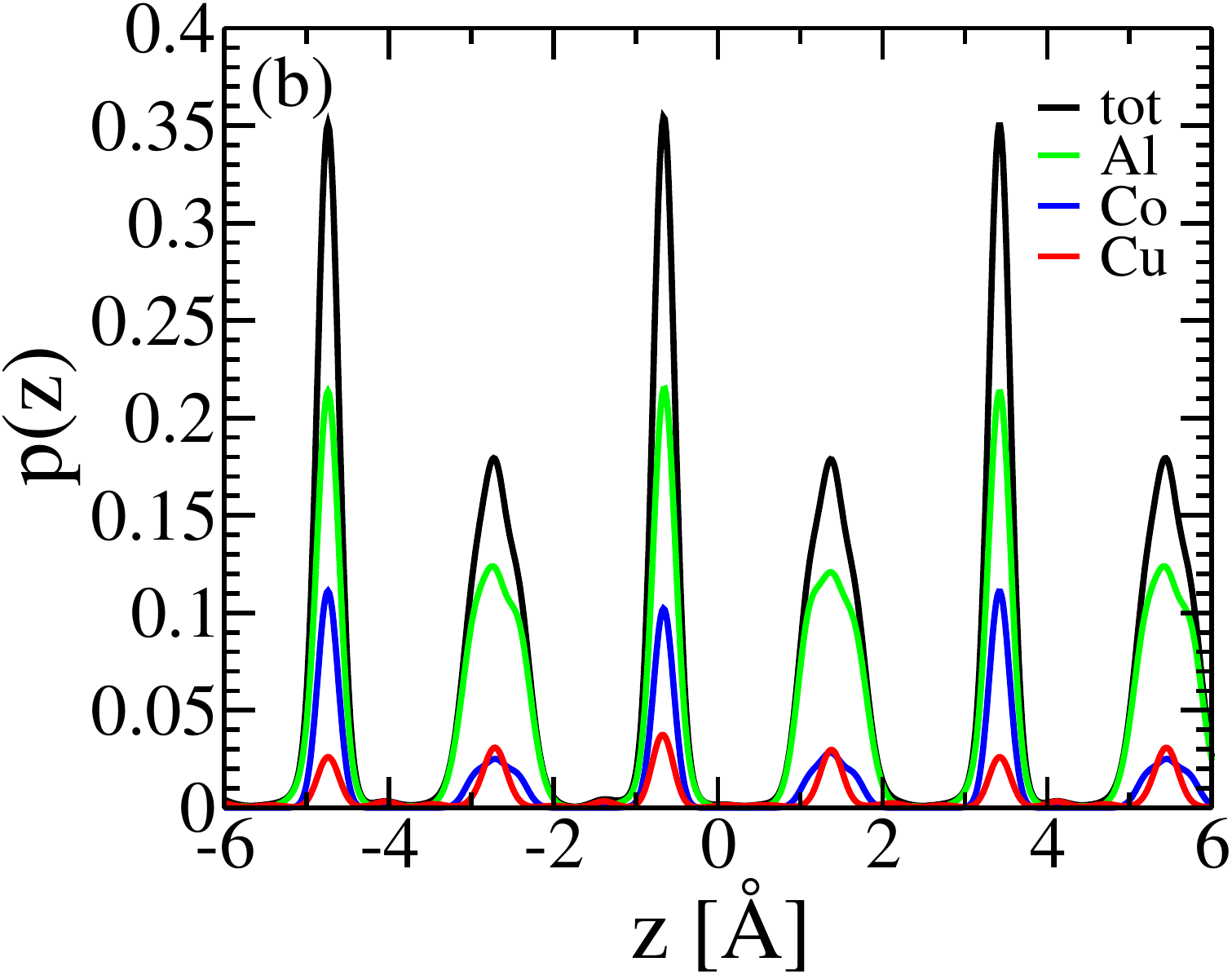}
  \caption{  \label{fig:zdist}
    Atomic distribution along $z$ direction in (a) Cu-rich and (b)
    Co-rich QCAs. Distributions of three species---Al, Co and Cu---are
    shown in green, blue and red, respectively, with the total in
    black. Positions of mid-layer Cu atoms are indicated by red arrows in (a).
}
\end{figure}

Adjacent 13~\AA~ clusters create overlap of the 20~\AA~ clusters that, in turn, can be decomposed into HBSD tilings, as shown in our simulated HAADF image (Fig.~\ref{fig:Cu-rich-HAADF}a). An approximately equal number of upward and downward clusters are found which explains the overall 10-fold symmetry of $d$-AlCoCu phase. The equivalence of the two layers is reflected in the apparent 2~\AA~ periodicity along the $c$-axis that can be observed in Fig.~\ref{fig:zdist}. However, the alternation of layer height with orientation creates a screw axis and converts the actual periodicity to 4~\AA. 

Different HAADF images are observed when projecting along the $x$ axis (perpendicular to the mirror planes) as shown in Fig.~\ref{fig:Cu-rich-HAADF}b and along the $y$ axis (parallel to the mirror planes) as shown in Fig.~\ref{fig:Cu-rich-HAADF}c. These directions are sometimes referred to, respectively, as ``between'' and ``along''~\cite{Rabson1991}. In Fig.~\ref{fig:Cu-rich-HAADF}b bright spots align, while in Fig.~\ref{fig:Cu-rich-HAADF}c they form a zig-zag with 4~\AA~ periodicity and the brightness alternates in columns forming an array of short and long intervals in a Fibonacci sequence which agrees with Ref.~\cite{Duguet2009}. All these features of our simulated HAADF images yield excellent agreement with experimental HAADF images in Ref.~\cite{Taniguchi2008,Yang2022}.

\subsubsection{Structure of Co-rich QCs}

\begin{figure}[htpb]
  \centering
  \includegraphics[width=.49\textwidth]{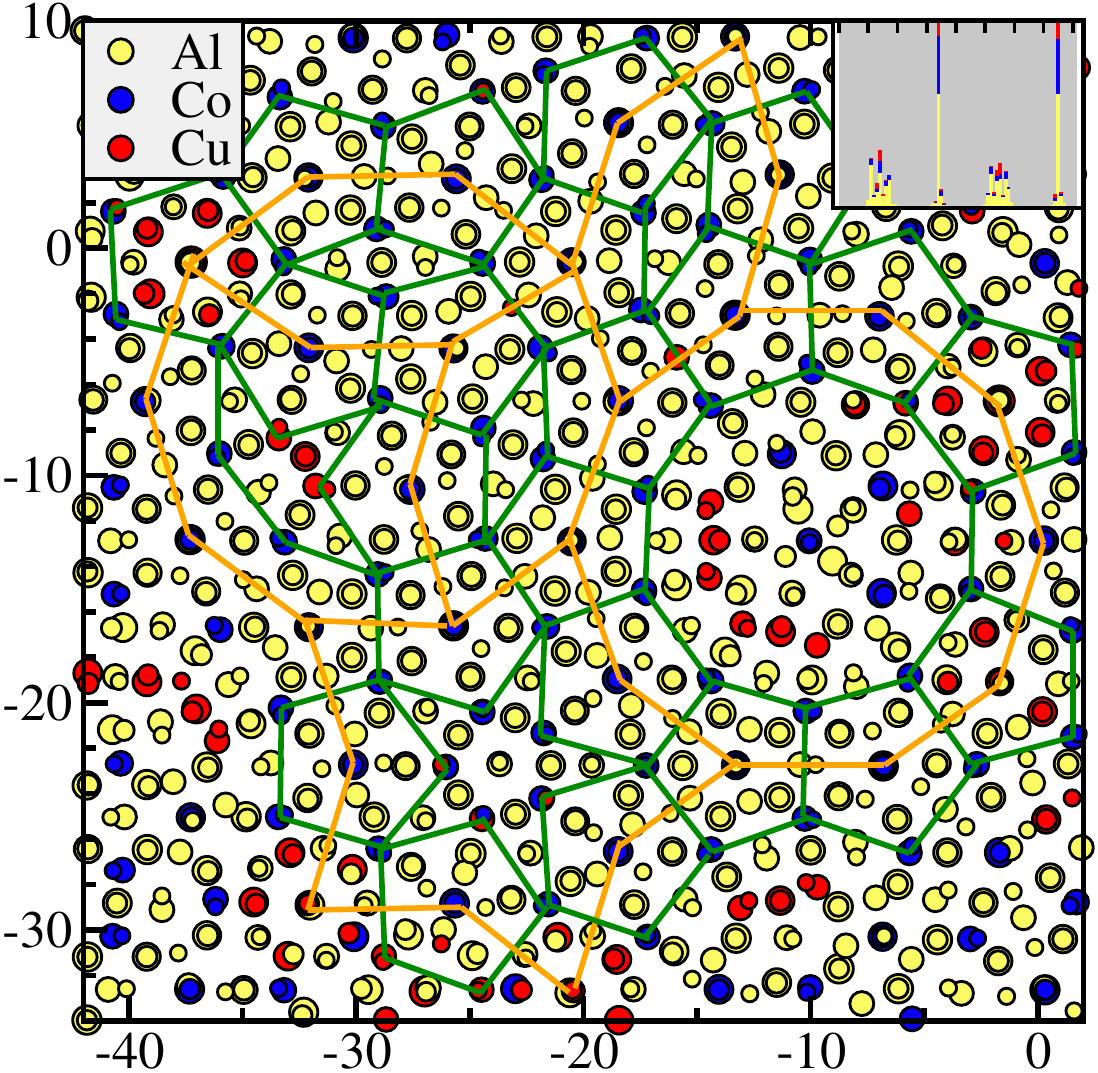}
  \caption{  \label{fig:Co-rich}
    HBSD tiling pattern (orange) and PB motifs (dark green) in a 8~\AA~ Al$_{1040}$Co$_{293}$Cu$_{142}$.}
\end{figure}

Complete Burkov-like clusters are less frequently seen in Co-rich structures (see  Fig.~\ref{fig:Co-rich}) owing to a shortage of Cu atoms. Rather, space is nearly filled with PB-like clusters. As we discuss in Sec~\ref{sec:mC102}, the PBs form locally 8~\AA~ periodic structures, alternating flat and puckered layers. However, PB equators and junctions may occur in the same layer, and this tends to average out the difference between the two flat layers. As a result, Co-rich QCAs exhibit 4~\AA~ periodicity, which agrees with experimental diffraction patterns~\cite{Grushko1993_LowCu,Grushko1993_HighCu,Nakayama2016}. Atomic distributions $\rho(z)$ of Co-rich QCAs (Fig.~\ref{fig:zdist}b) show 4~\AA~ periodicity due to alternating flat and puckered layers. This oscillation is most pronounced for Co atoms, which are more prevalent in flat layers than in puckered ones. Connecting the centers of two adjacent PBs of opposite orientation as shown in Fig.~\ref{fig:Co-rich}, an HBSD tiling pattern is observed. In our Co-rich QCA, all star tiles take the same orientation creating five-fold symmetry~\cite{Gu2006}. In contrast, our Cu-rich QCA (Fig.~\ref{fig:Cu-rich-HAADF}) display both orientations, restoring the ten-fold symmetry.

\section{Quasicrystal at finite temperature}

To study the quasicrystal's stability at finite temperature, we investigate the 83-atom quasicrystal approximant
Al$_{53}$Co$_{15}$Cu$_{15}$ ($Q8$) and its competing phases---$S4$
($\tau$-Al$_{12}$Co$_{3}$Cu$_{5}$.hP5), $S8$ (Al$_{72}$Co$_{24}$Cu$_6$.mC102) and $S11$
(Al$_9$Co$_2$.mP22). Its relative free energy is
\begin{equation}
  \Delta G=G(Q8)-x_1 G(S4)-x_2 G(S8)- x_3 G(S11)
\end{equation}
where
\begin{align}
  x_1=0.63834,\ x_2=0.35933,\ \hbox{and}\ x_3= 0.00233.\nonumber
\end{align}
Evidently, the $\tau$ phase (hP5) is the most important competitor, followed by the Co-rich QCA (mC102). The Gibbs free energy
\begin{equation}
  \label{eq:G}
  G = \Delta H + G^v + G^e + G^a
\end{equation}
contains the relaxed enthalpy $\Delta H$ calculated from first principles as in Eq.~(\ref{eq:dH}), harmonic vibrational and electronic free energies ($G^v$ and $G^e$), and the anharmonic free energy ($G^a$, Eq.~(\ref{eq:Fa})) calculated from EOPP simulations. Among these contributions, the harmonic vibrational free energy is the most important at low temperatures, with the anharmonic part becoming increasig important at high temperatures.

\begin{figure}[htpb]
    \centering
    \includegraphics[width=.45\textwidth]{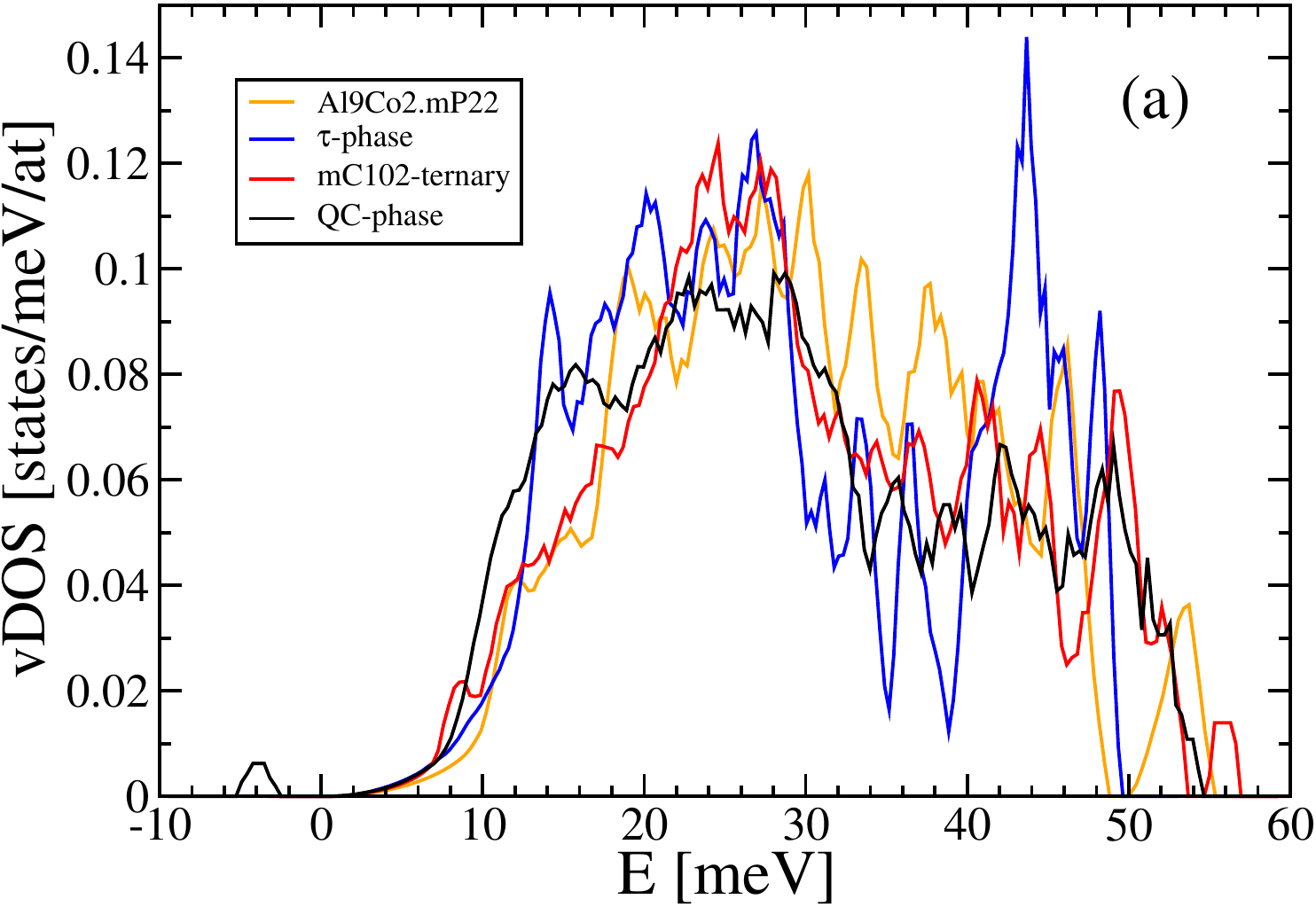}
  \includegraphics[width=.45\textwidth]{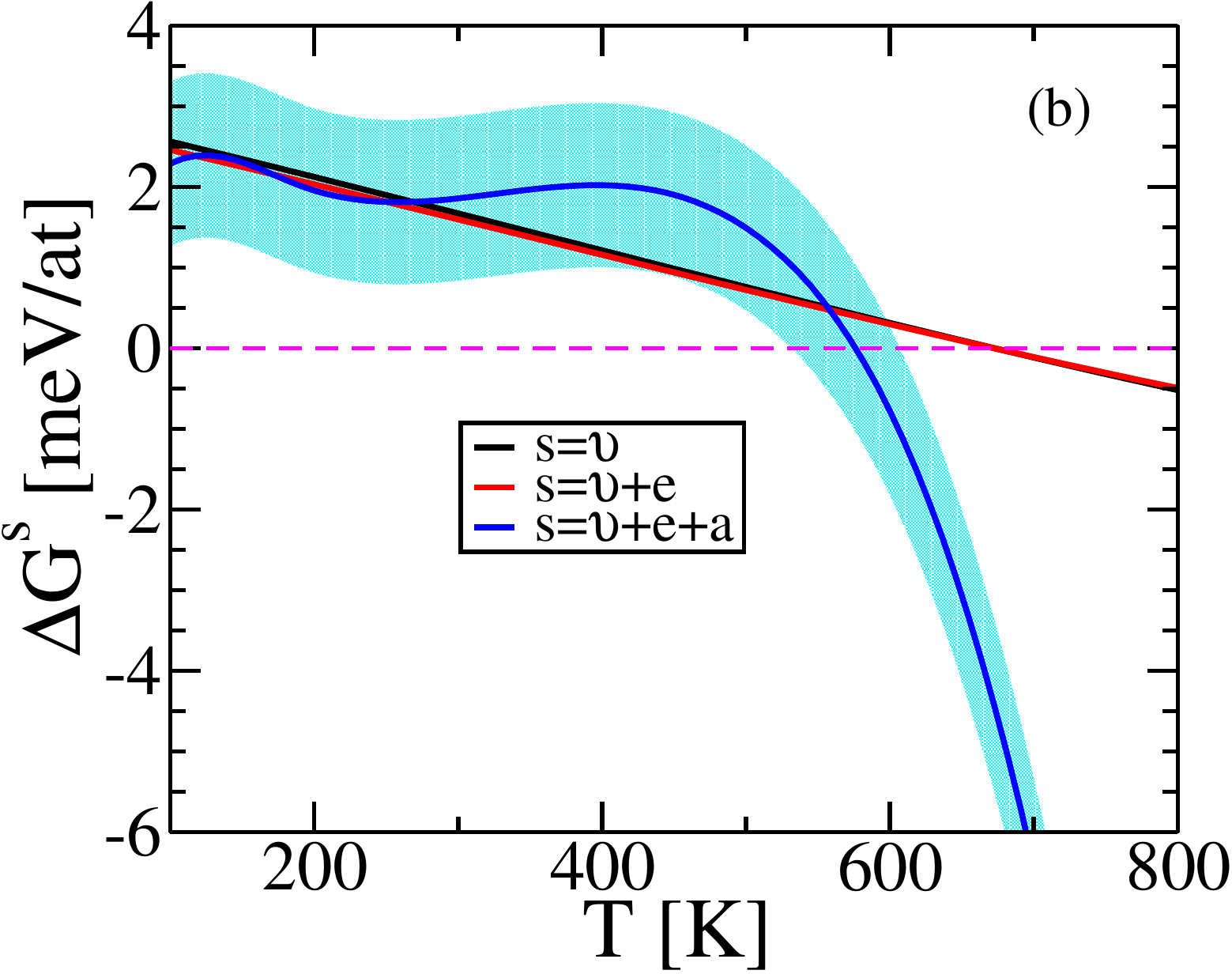}
    \caption{(a) Vibrational density of states of the 83-atom QCA and its competing phases; (b) Relative Gibbs free energy of the 83-atom QCA including $G^v$, $G^e$, and $G^a$. Shading indicates estimated uncertainties.}
    \label{fig:vDOS}
  \end{figure}

Vibrational densities of states are shown in Fig.~\ref{fig:vDOS}a. Q8 has an excess of low frequency vibrational modes around 10 meV relative to the competing phases. These modes will reduce its quantum zero point energy at $T=0$K, and substantially reduce its vibrational free energy as temperature rises.  Phonopy claimed a small number of imaginary modes in the QCA that are omitted from $G^v$, and these would have further stabilized the QCA (see App.~\ref{app:EOPP}). The relative Gibbs free energies are shown in Fig.~\ref{fig:vDOS}b, where we see that $G^v$ alone is sufficient to stabilize Q8 above 700K.

\begin{figure}[htpb]
  \centering
  \includegraphics[width=.23\textwidth]{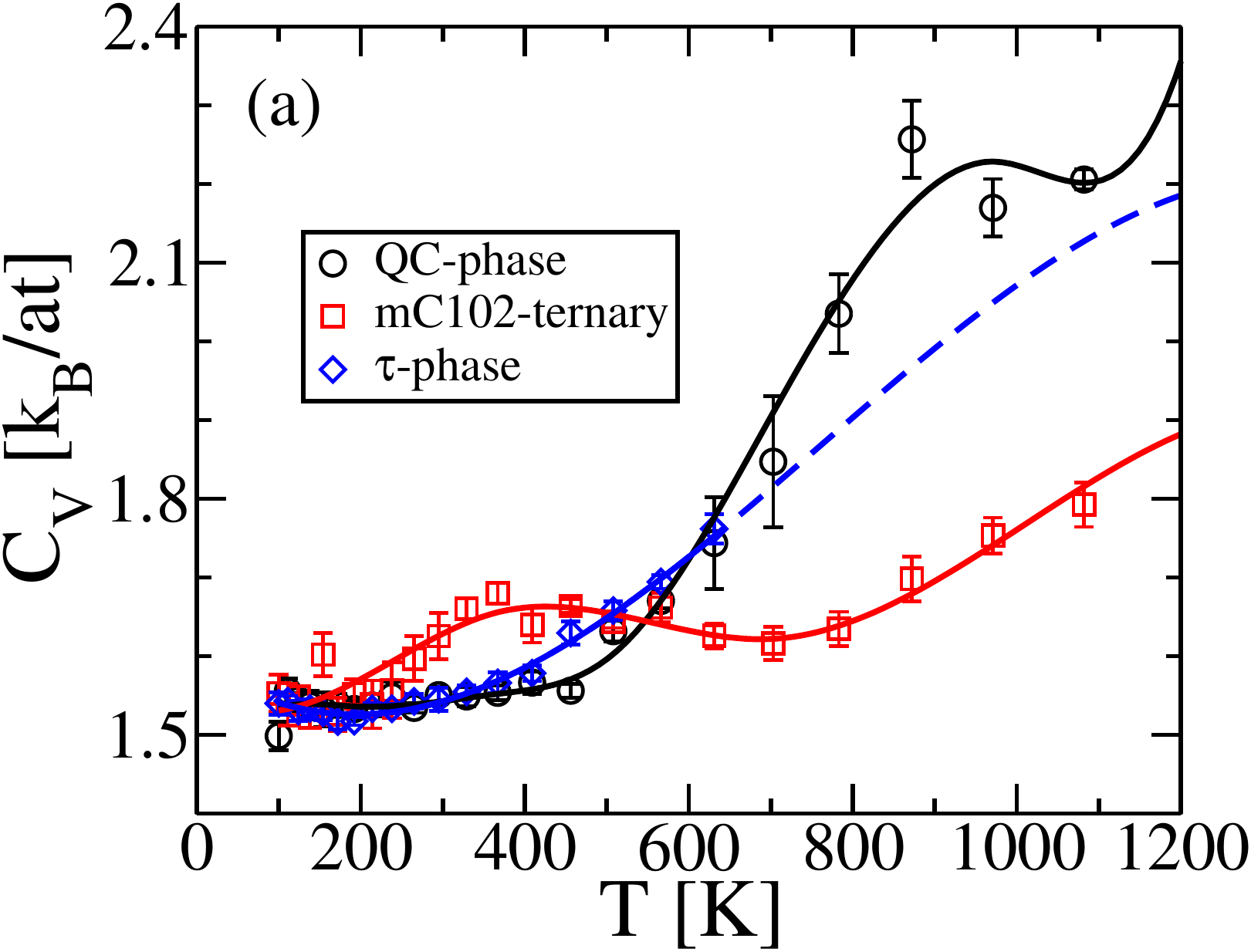}
  \includegraphics[width=.23\textwidth]{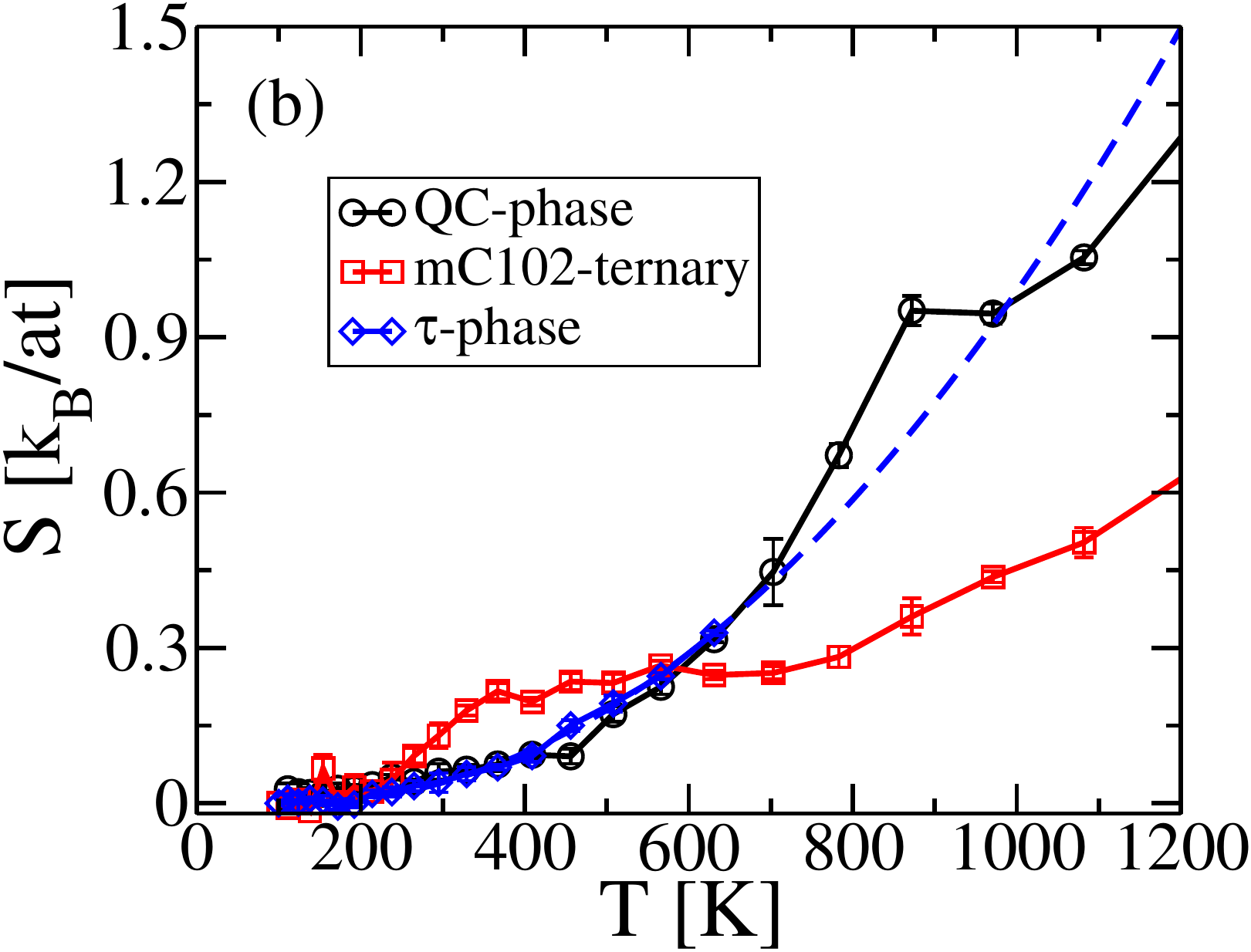}\\
  \includegraphics[width=.23\textwidth]{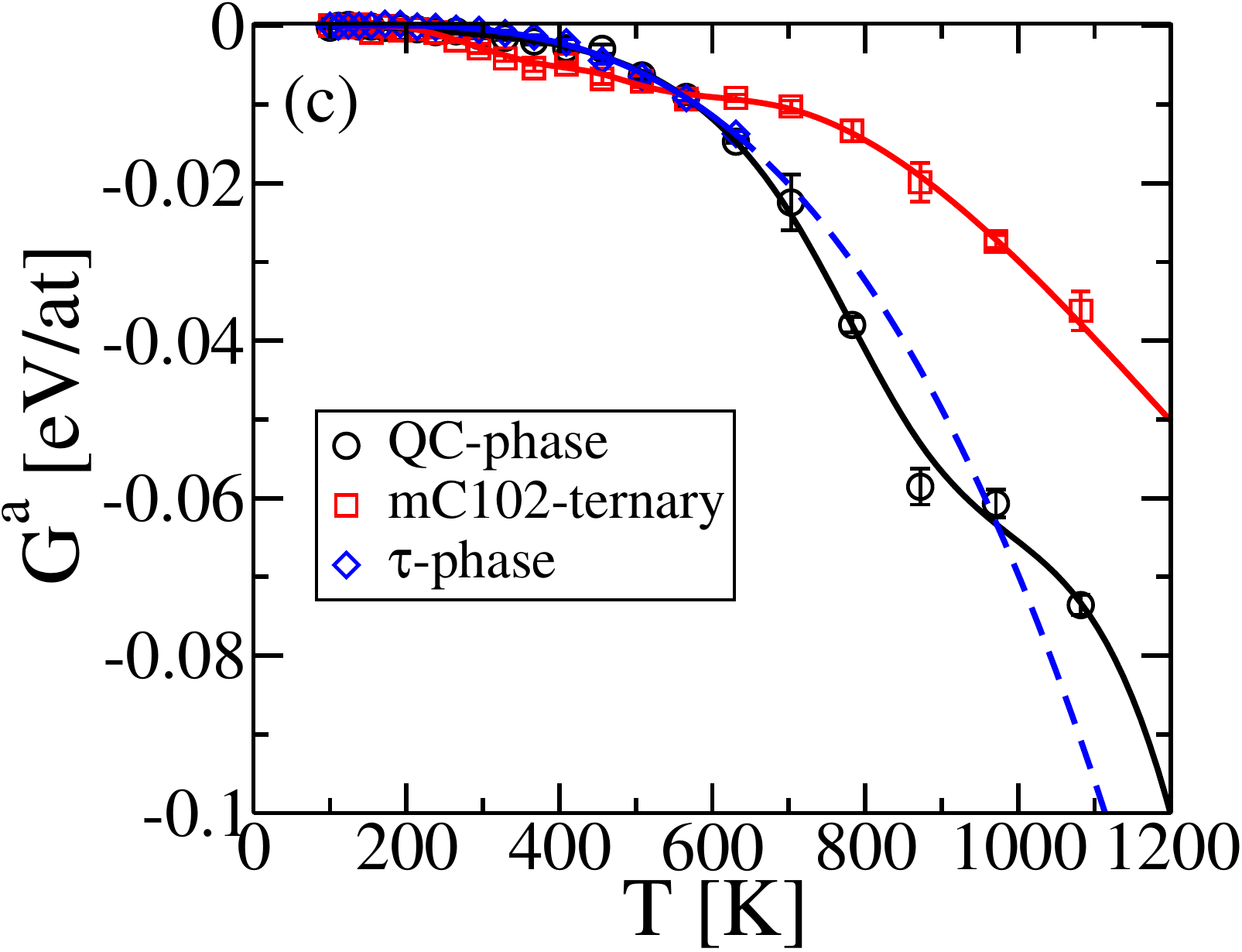}
  \includegraphics[width=.23\textwidth]{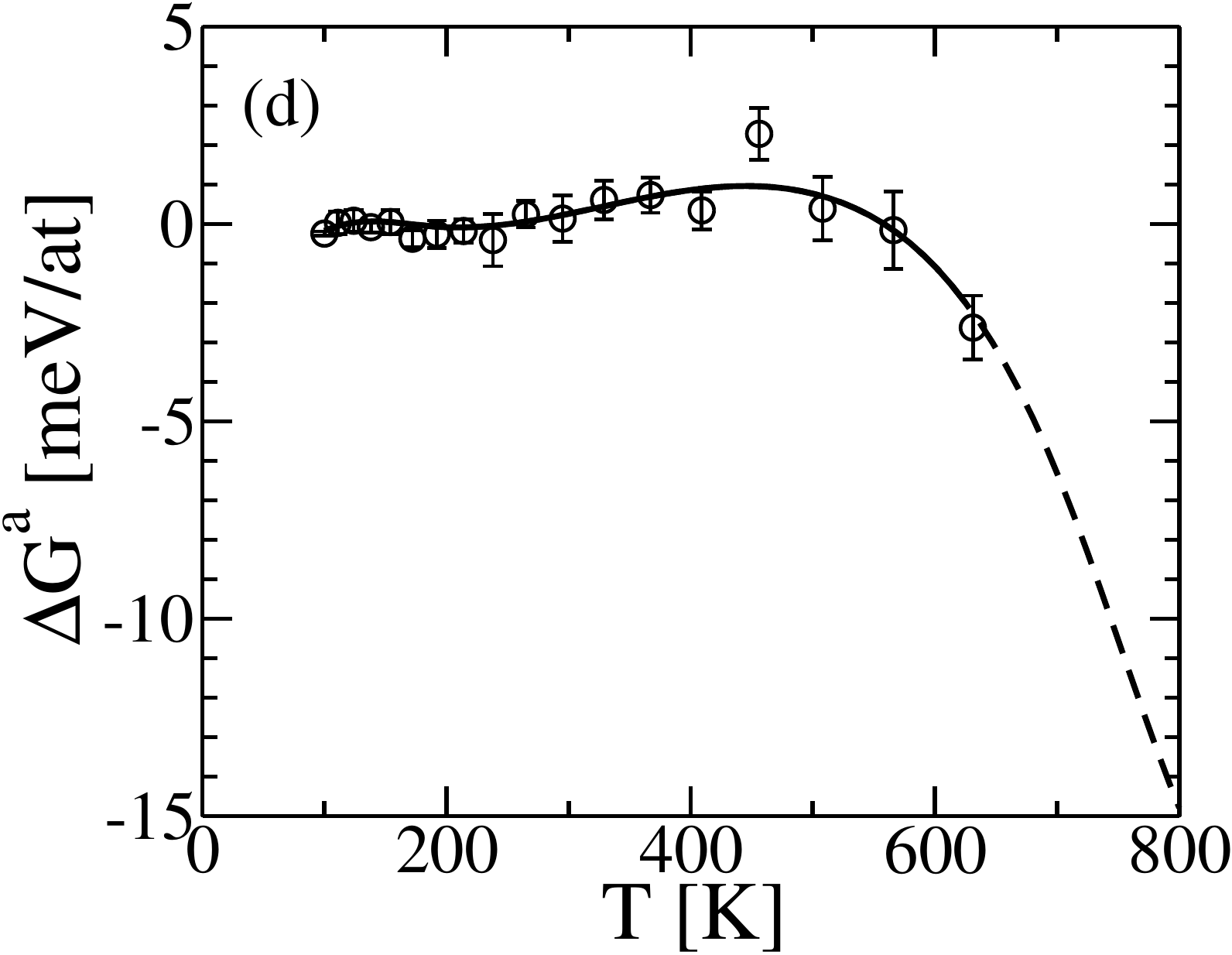}
  \caption{\label{fig:5}Calculated thermodynamic properties of the quasicrystal approximant Q8 and its principal  competing phases. (a) Heat capacities $c_v$. (b) Entropies $S$. (c) Anharmonic free energies $G^a$. (d) Anharmonic free energy difference $\Delta G^a$. Error bars are estimated from three independent runs. Data points are interpolated with cubic splines weighted by error inverse. For the $\tau$ phase, they are extrapolated to high temperatures as shown in dashed blue lines (see Appendix~\ref{app:tau}).}
  \end{figure}

  To assess the anharmonic effects, Fig.~\ref{fig:5}a compares the heat capacities of Q8 (in a 166-atom 16~\AA~ $1\times 1\times 2$ supercell) with its competing phases, mC102 and $\tau_3$ (in a 160-atom $4\times 4\times 2$ supercell). mC102 has the highest heat capacity below $500$K because intra-layer Al-Cu swaps are low in energy. Beyond this broad peak, inter-layer swaps at high energy cost reduce the heat capacity relative to the quasicrystal and $\tau_3$. The quasicrystal heat capacity grows rapidly beyond 600K due to the onset of interlayer phason disorder (see App.~\ref{app:phason}). Fig.~\ref{fig:5}b presents the integrated anharmonic entropy $S^a$. As expected based on heat capacities, mC102 has a large excess entropy at low temperature while the quasicrystal takes the lead for $T>700$K. We further calculate anharmonic free energies $G^a$ in Fig.~\ref{fig:5}c and relative anharmonic free energy $\Delta G^a$ of quasicrystal respect to its competing phases in Fig.~\ref{fig:5}d. The relative free energy $\Delta G^a$ is positive for $T<600$K but falls off rapidly when temperature increases which tends to stabilize the quasicrystal phase.

Finally, we add up all relative free energy contributions---enthalpy $\Delta H$, vibrational free energy $\Delta G^v$, electronic free energy $\Delta G^e$ and anharmonic free energy $\Delta G^a$--- to predict the stabilization temperature of Q8 (Fig.~\ref{fig:vDOS}b). The vibrational contribution is capable of stabilizing the quasicrystal phase at around $T=700K$. The electronic contribution has a negligible correction to the total free energy. Meanwhile the anharmonic contribution lowers the temperature down to about $590$K. Because of uncertainties in our simulations and calculations, plausible stabilization temperatures for Q8 lie within $530$-$610$K. Similar calculations are performed for a 217-atom $a_4$ (L0) and a 569-atom $a_5$ (L19) approximants (See Appendix~\ref{app:size}). Heat capacities $c_v$, entropies $S$, anharmonic free energies $G^a$  and relative anharmonic free energies $\Delta G^a$ are then calculated and compared with those of Q8 in Fig~\ref{fig:thermo_size}. Transition temperatures are predicted to be 590K, 620K and 780K with larger uncertainties, and with higher transition temperatures due to their increasing enthalpies $\Delta H$.

At composition Al$_{65}$Co$_{17.5}$Cu$_{17.5}$, experiment finds that the decagonal phase is stable in the temperature range between 973 and 1350K~\cite{Dong_1991}. At lower temperatures it transforms into a microcrystalline approximant structure. Congruent melting was claimed at 1350K, although more recently peritectic melting was reported at $1040^\circ$C~\cite{BOGDANOWICZ2002255}.  At composition Al$_{65}$Co$_{20}$Cu$_{15}$ the quasicrystal does not transform after annealing at $550$$^\circ$C for 32 days\cite{He1991}. These experiments agree with our free energy calculations that predict stability of the quasicrystal at elevated tempertures.

\section{Conclusion}

We perform replica exchange MD simulations for $d$-AlCoCu near composition Al$_{65}$Co$_{17.5}$Cu$_{17.5}$ using EOPP to optimize structure models, followed by accurate first-principle relaxation energies.  Because of a variety of competing phases near this composition, we find the dQC is energetically slightly unstable, with several 83-atom QCAs having $\Delta E$ within 3-4 meV/atom of the convex hull. Enthalpies increase beyond 10meV/atom for larger QCAs. We compute the entropy by integrating heat capacity from our MCMD simulations, showing that the QC has relatively high entropy at high temperature.  Enhanced high temperature heat capacities have been observed experimentally in $d$-AlCoCu and $i$-AlMnPd~\cite{Edagawa2000} and also in $i$-AlPdRu~\cite{Kimura2023}. By calculating the Gibbs free energy, we predict the quasicrystal to be entropically stabilized at elevated temperatures. This conclusion is consistent with experimental evidence that the quasicrystal transforms into crystalline states at low temperatures~\cite{Hiraga1991}.

We extend our study up to QCAs containing more than 1400 atoms over a wide range of composition from Cu-rich QCA to Co-rich. The Cu-rich QCA shows a 4~\AA~ periodicity with a $10_5$ screw axis that is caused by a 36$^\circ$ rotation between adjacent layers. While the Burkov model predicts a ten-fold cluster, we observe a locally five-fold cluster with two possible orientations. The two occur with equal frequency in Cu-rich structures, explaining the global ten-fold symmetry. The Co-rich structures are dominated by locally 8~\AA~ PB motifs. It also shows a global 4~\AA~ periodicity caused by mixing of heights of the PB equators. Our model predicts five-fold symmetry in the Co-rich limit.

Our thermodynamic calculations show that quasicrystal approximants are stabilized at elevated temperatures against competing crystalline phases. Vibrational free energy alone would be sufficient to stabilize Q8 above 700K, but this does not explain the occurence of the true quasicrystal. The random tiling hypothesis~\cite{HenleyRT} provides a possible mechanism to explain the emergence of true quasiperiodicity. In this picture, a random tiling exhibits long-range quasiperiodicity with maximal probability; the ensemble of random tilings is sampled through a sequence of phason flips. Our simulations indeed exhibit phason flips, recognized as swaps of nearby Co/Cu atoms. The hypothesis of entropic stabilization~\cite{BinaryLJ,WidomRT} suggests that the random tiling entropy provides the margin needed to stabilize the QC against its competing phases. Here we estimated the phason entropy to provide approximately 7 meV/atom of stability at 1000K, an amount similar in magnitude to the free energy differences among competing phases.

\subsection{A high entropy quasicrystal?}
High entropy alloys~\cite{Yeh04_1,Yeh04_2,Cantor04} refer to multi-principle-element compounds that exhibit strong chemical substitutional disorder among their constituent elements. If they contain $N$ elements, their substitutional entropy is expected to be close to the ideal entropy of mixing,
\begin{equation}
  \label{eq:Sideal}
  S_{\rm Ideal}/\kB=-\sum_\alpha x_\alpha\ln{x_\alpha}\sim\ln{N}.
\end{equation}
For our Al-rich dQC compositions we find $S_{\rm Ideal}\approx 0.89\kB$, which is an order of magnitude greater than our simulations reveal. In this sense, even though the quasicrystal appears to be entropically stabilized it does not qualify as a high entropy alloy. The reason is that the chemical substitutions that occur are limited to specific subsets of atomic positions, as illustrated in Fig.~\ref{fig:Cu-rich-atomic}b. Note also that Co substitutes with Cu, and Al substitutes with Cu, but Co does not directly substitute with Al. An equiatomic high entropy quasicrystal~\cite{ZHANG2022} was found to be metastable.

It is natural to ask if a genuinely high entropy quasicrystal might exist. Given the need to preserve a diverse set of atomic site types ({\em e.g.} the Co atoms often center Al pentagons) we adopt the strategy of seeking substitutions within specific site classes, defined by the nominal atomic species. This is analgous to the strategy used in high entropy ceramics, where substitutions occur only on cation sites~\cite{Oses2020}. Further, since the electronic pseudogap is important for stabilizing the quasicrystal, we propose only isoelectronic substitutions, an approach that has frequently been used to predict new quasicrystals~\cite{ICQ15}.

Taking advantage of the ability to substitute Co and Cu, we randomly substituted Ni in the Q8 structure ($\Delta H=3.0$ meV/atom above the convex hull), yielding Al$_{53}$(Co,Cu,Ni)$_{30}$ that lay $\Delta H=12$ meV/atom above the convex hull. Substituting Co with (Co,Ir,Rh), within the same periodic table column resulted in Al$_{53}$Cu$_{15}$(Co,Ir,Rh)$_{14}$ at $\Delta H=10$ meV/atom. Subsituting Ga with (Al,Ga) yielded (Al,Ga)$_{53}$Co$_{15}$Cu$_{15}$ at $\Delta H=28$ meV/atom. In each case the rise in energy can be compensated by the corresponding entropy increase at temperatures in the range 400-800K. Substitution into the $\tau_3$ or B2 phases will also increase their entropies, but the increase will be proportionally less than in the quasicrystal owing to the high substitutional entropy those phases posess even in the Al-Co-Cu ternary. In fact, rapidly quenched $d$-AlCoCuNi has already bee reported~\cite{YADAV2007}.

Similar strategies would likely work for the Al-Cu-Fe icosahedral phase, replacing Fe with (Mn,Fe,Co), or with (Fe,Ru,Os), and possibly also Al with Ga.

\section{Acknowledgements}
MM is thankful for the support from the Slovak Grant Agency  VEGA (No. 2/0144/21) and APVV (No. 20-0124, No. 19-0369). YH and MW acknowledge support of the US Department of Energy grant DE-SC0014506. This research also used the resources of the National Energy Research Scientific Computing Center (NERSC), a US Department of Energy Office of Science User Facility operated under contract number DE-AC02-05CH11231 using NERSC award BES-ERCAP24744. MM and MW acknowledge the hospitality of Michael Engel and the Friedrich Alexander University of Erlangen-Nurenberg.

\begin{appendix}
  \renewcommand\thefigure{\thesection.\arabic{figure}}
  \renewcommand\thetable{\thesection.\roman{table}}

\section{More about empirical oscillating pair potential}
\label{app:EOPP}
\setcounter{figure}{0}

\begin{figure}[htpb]
  \centering
  \includegraphics[width=.23\textwidth]{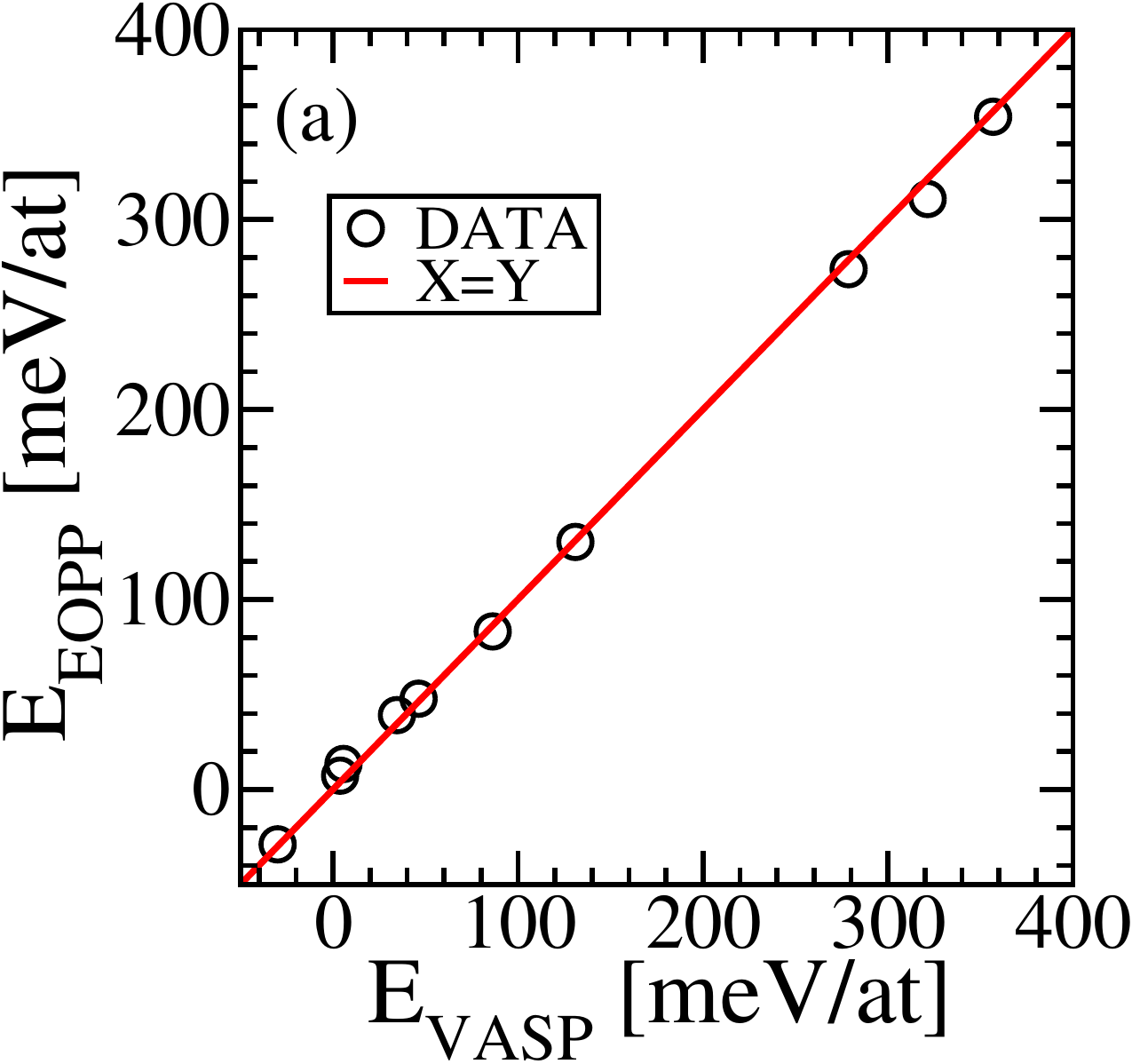}
  \includegraphics[width=.23\textwidth]{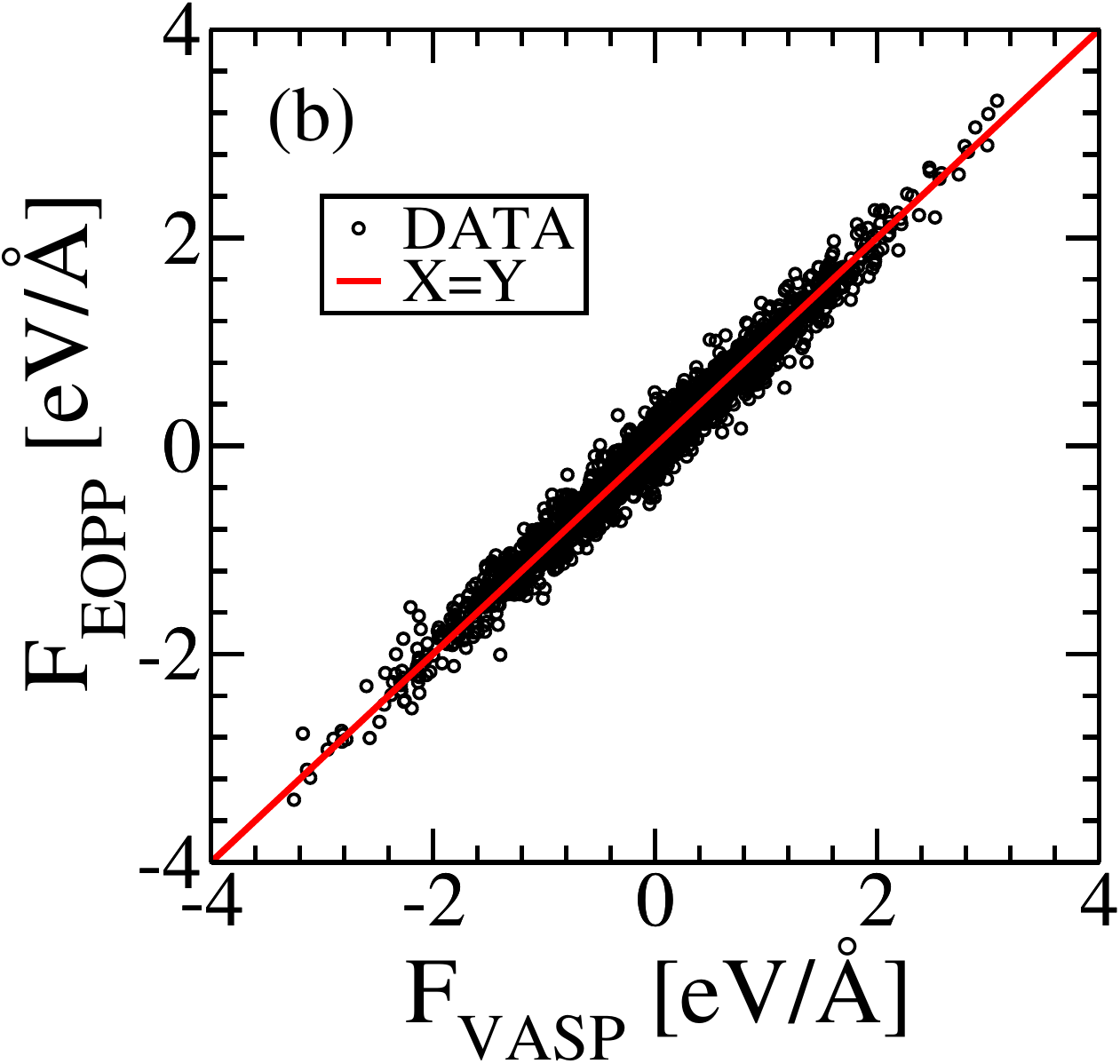}
  \caption{  \label{fig:ppfit}
    Comparison of VASP-calculated and pair potential fitted energies (a) and forces (b).}
\end{figure}

The pair potential is fitted to relative energies and forces of 11
solid structures and 2 liquid systems. Specifically, we consider fully relaxed Al.cF4, Al13Co4.oP102, Al2Cu.cF12, Al2Cu.tI12, Al3Co.oI96, Al6CoCu3.hP5, Al7CoCu2.tP40, Al9Co2.mP22, Al13Co4.q01, an H2S-tiling of Al-Co-Cu, AlCu.tP40-221, and liquid Al at 600K and liquid AlCu at temperature 1000K. Relative energies are calculated respect to Al.cF4, Al3Co.oI96 and Al2Cu.cF12. Parity plots comparing VASP-calculated and fitted values are shown in Fig.~\ref{fig:ppfit}. TABLE~\ref{tab:2} lists parameters for all 6 pairs.

\begin{table}[htpb]
  \centering
  \begin{ruledtabular}
    \begin{tabular}{ccccccc}
      Pair&$C_1$ &$\eta_1$&$C_2$ &$\eta_2$&$k$&$\phi$\\
      -&[eV$\cdot$\AA$^{\eta_1}$]&-&[eV$\cdot$\AA$^{\eta_2}$]&-&[\AA$^{-1}$]&-\\
      \hline
      Al-Al &771.6200&  8.3309& -2.4184&  4.3472&  3.8752&-10.6377\\
      Co-Al &314.8235&  8.8841&  3.6873&  3.2567&  3.1113& -4.4691\\
      Co-Co &226.0575&  6.3040&  1.5747&  2.5289&  2.9913& -3.4225\\
      Cu-Al &164.5061&  7.1380& -2.6665&  3.5251&  3.1879& -7.8115\\
      Cu-Co &229.5015&  7.1213& -2.0137&  2.8701&  3.0729& -7.0260\\
      Cu-Cu &304.2326&  7.8303& -0.9824&  2.7515&  3.1140& -6.9253
    \end{tabular}
  \end{ruledtabular}
  \caption{\label{tab:2} Fitted parameters for Al-Co-Cu.}
\end{table}

\begin{figure}[htpb]
  \centering
  \includegraphics[width=.23\textwidth]{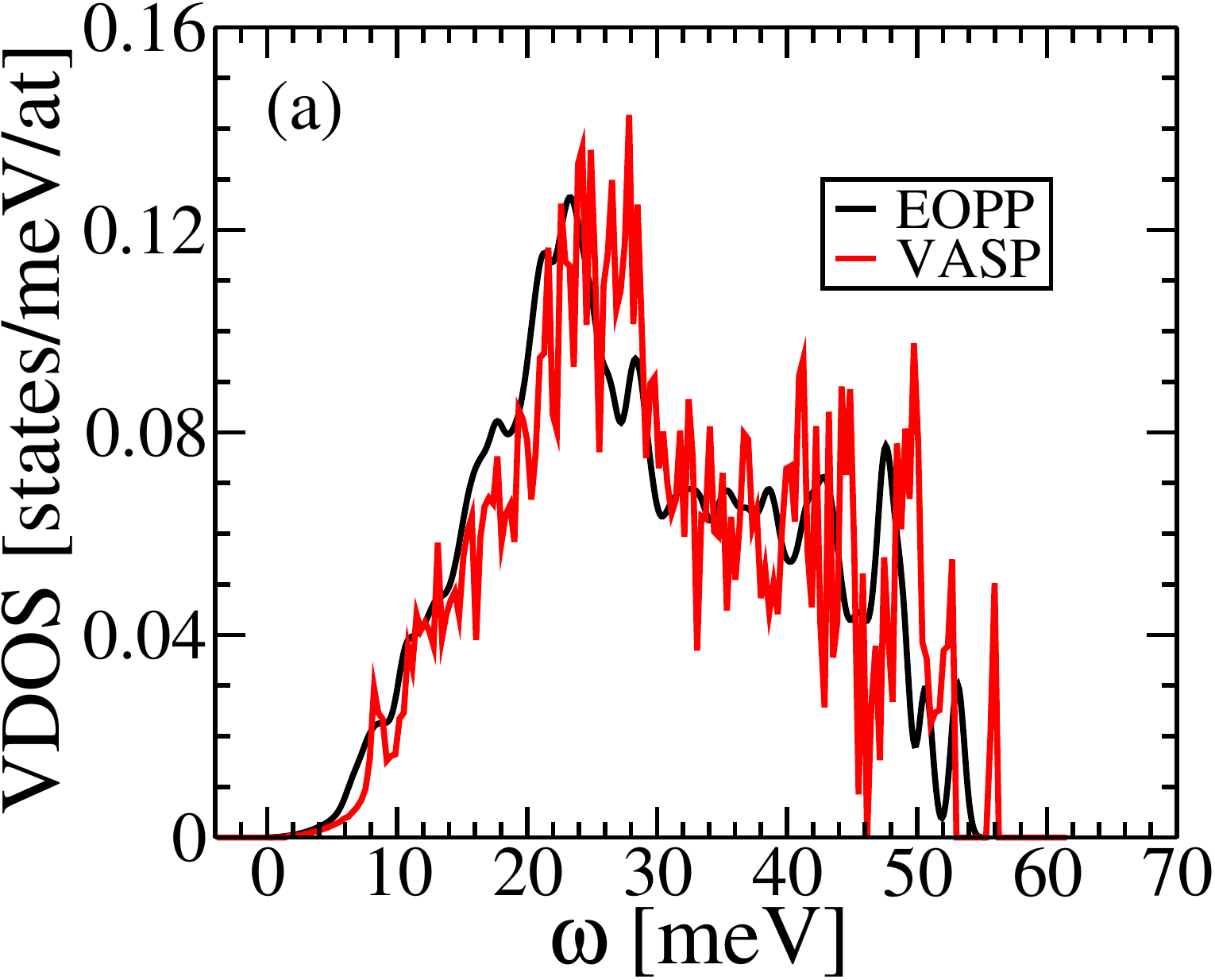}
  \includegraphics[width=.23\textwidth]{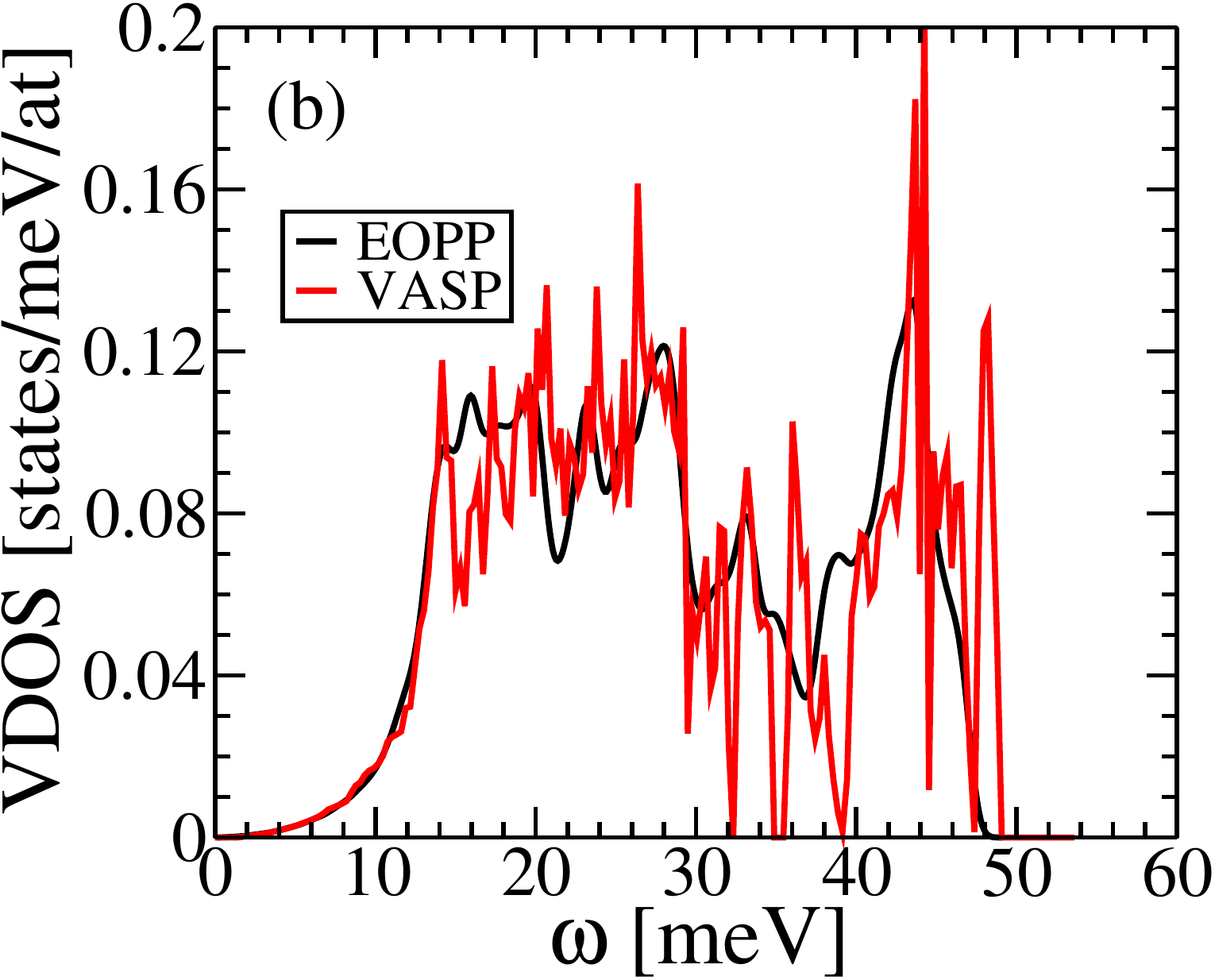}
  \\
  \includegraphics[width=.23\textwidth]{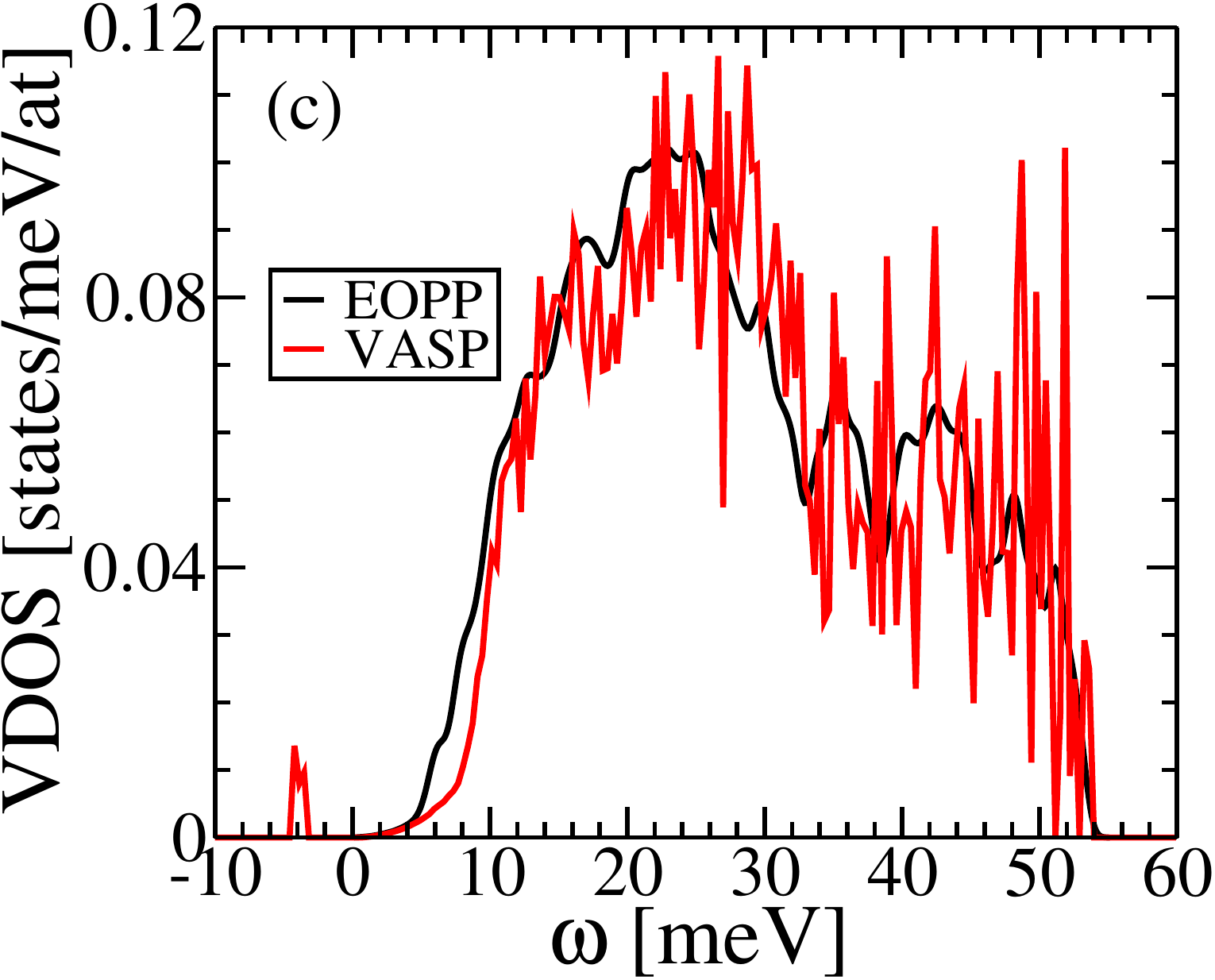}
  \includegraphics[width=.23\textwidth]{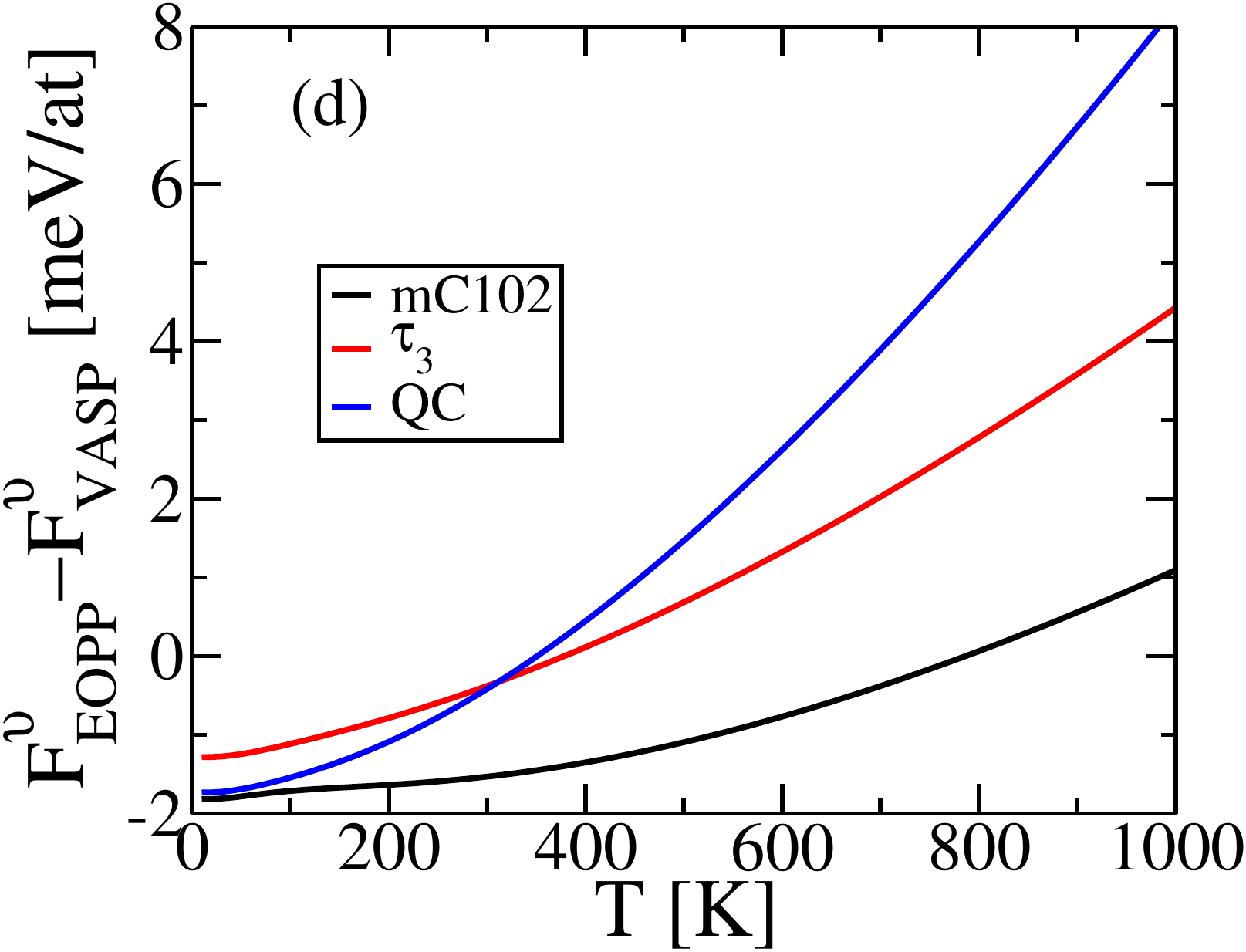}
  \caption{  \label{fig:12}
    Comparison of vibrational density of states with phonopy and    pair potential for three structures: (a) mC102, (b)  hP5 ($\tau_3$) and (c) quasicrystal (Q8). (d)~Differences in calculated    vibrational free energy. }
  \end{figure}

Although the vibrational free energies reported in the main text were calculated using VASP, full DFT calculation of vibrational free energy for large QCA is prohibitive, so we wish to apply EOPP instead. Fig.~\ref{fig:12}d compares the vDOS between the two methods for small systems, showing that the overall shapes and ranges match well. For Q8, EOPP solves the problem of imaginary modes. The two methods yield free energies that lie within a few meV/atom of each other.
% The $\tau$-phase is a vacant structure so that its vibrational free
% energy is \cite{Wurschum1994} 

\section{Convex hull - closeup}
\label{app:Hull-closeup}
\setcounter{figure}{0}

\begin{figure}[htpb]
  \centering
  \includegraphics[width=.4\textwidth]{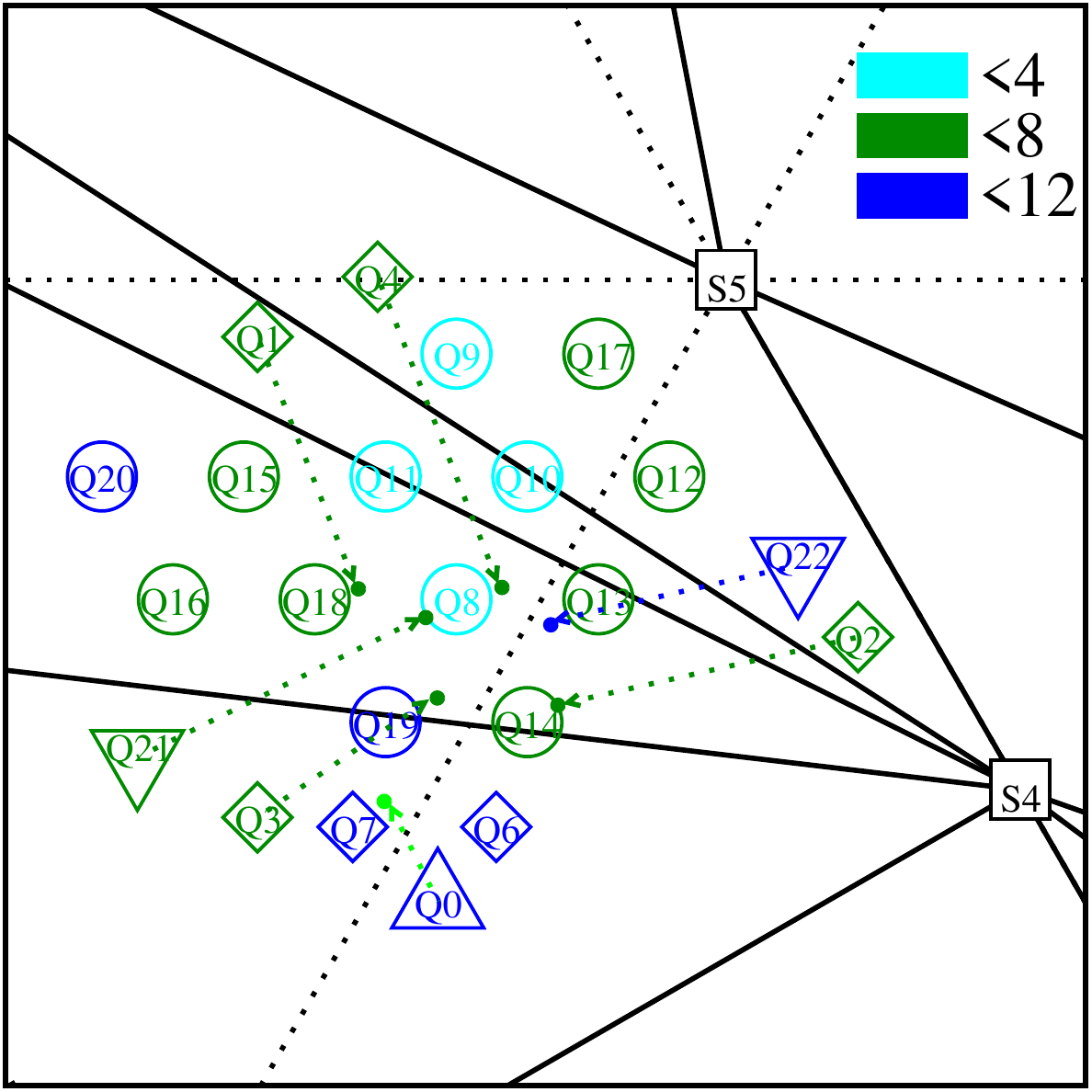}
  \caption{  \label{fig:Hull-closeup}
    Convex hull of Al-Co-Cu enthalpies showing stable phases in black. Structure information is summarized in Tab.~\ref{tab:Hull} and Tab~\ref{tab:Hull-closeup}. Circles=83 up triangles=81 down triangles=84  diamonds=82 colors are cyan$<$4 meV/atom, green$<$8, dark blue$<$12.}
\end{figure}

A wide range of densities and compositions of quasicrystals near Al$_{65}$Co$_{17.5}$Cu$_{17.5}$ with $a=7.185$\AA~ and $c=4.080$\AA\cite{Dong_1991}  have been investigated with pair potential MCMD replica exchange simulation. Their composition map is shown in Fig.~\ref{fig:Hull-closeup} and relative energies are summarized in Tab.~\ref{tab:Hull-closeup}. The best $a_3$ approximants have 81-84 atoms/cell with Al concentration less than 84 at.\% and nearly equal Co and Cu content.
%$\rho=0.07134$ atom/\AA$^3$\ at Al$_{53}$Co$_{14}$Cu$_{16}$, Al$_{52}$Co$_{16}$Cu$_{15}$, Al$_{52}$Co$_{15}$Cu$_{16}$ and Al$_{53}$Co$_{15}$Cu$_{15}$.
First-principles calculations predict relative energies $\Delta E$ within $3-4$ meV respect to their competing phases. 82-atom approximant Q1 and 84-atom approximant Q21 also have promising relative energies $\Delta E$ near $6$ meV. We provide a selection of approximant structure coordinates and their competing phases in the Supplemental Material.

\begin{figure}[hptb]
  \centering
  \includegraphics[width=.45\textwidth]{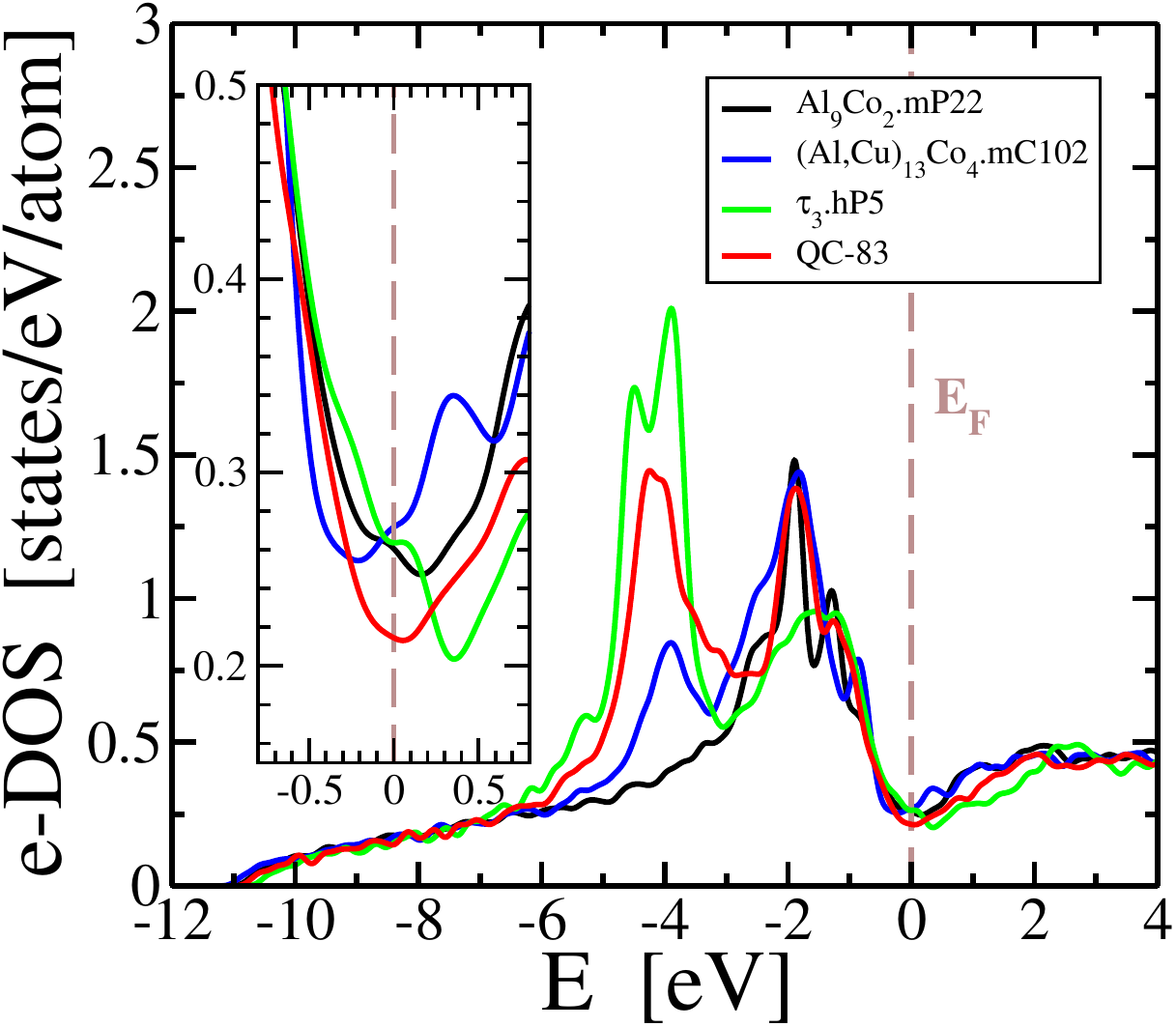}  \includegraphics[width=.45\textwidth]{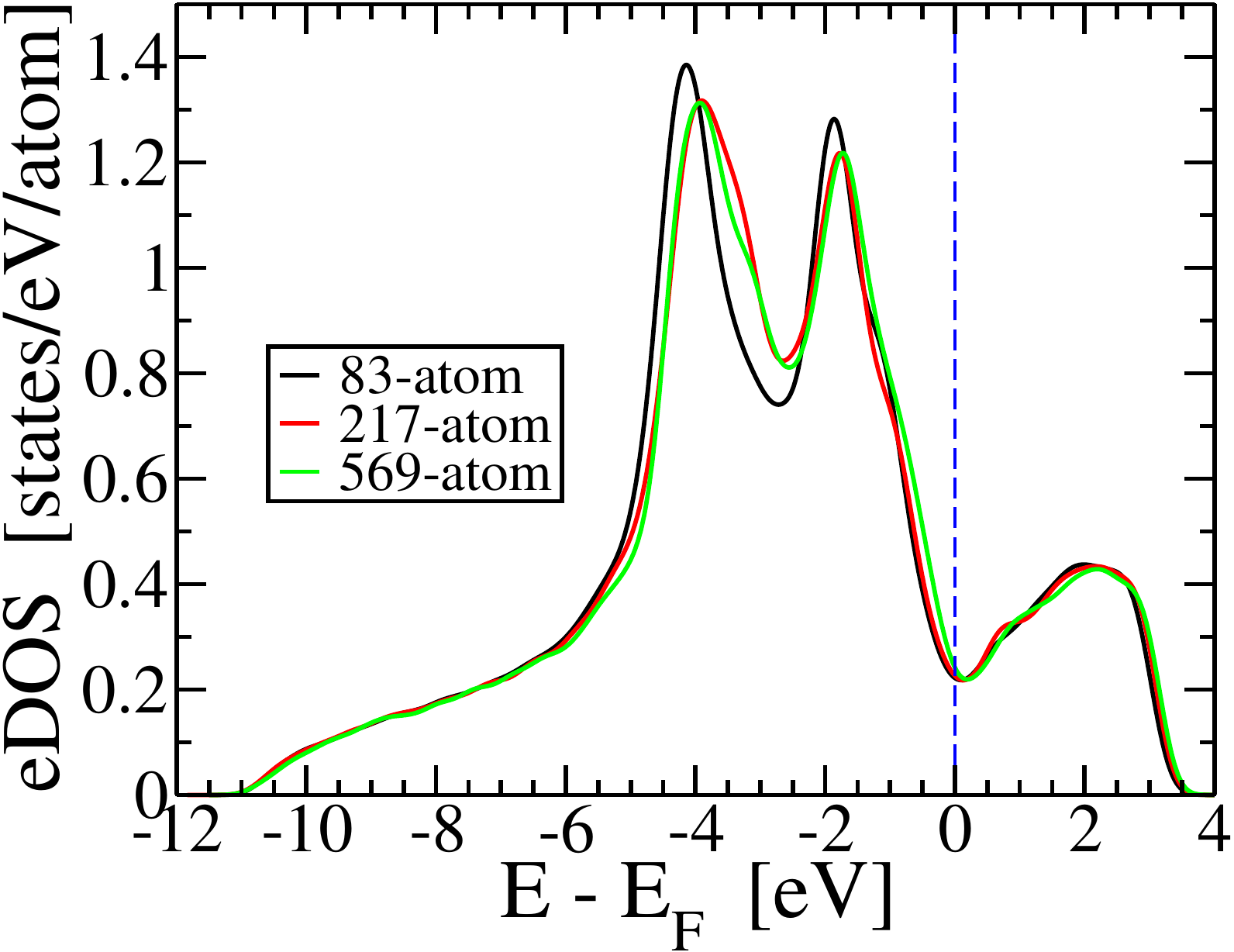}
  \caption{\label{fig:eDOS} (a) Comparison of electronic densities of states (eDOS) of Q8 with its competing crystal phases. (b)    Electronic density of states (eDOS) of different QCAs. Calculations have been done using tetrahedron integration of the Brillouin zone followed by Gaussian smearing of 0.1 eV. 83=Q8, 217=L0; 569=L19.}
  %QC_217-atom.dos: symbolic link to /home/hyhy123/alloy.pbe/AlCoCu/quasi/frommarek/80-atom-tau/217-atom.equal/Al138Co39Cu40/392/dos
%QC_569-atom.dos: symbolic link to /home/hyhy123/alloy.pbe/AlCoCu/quasi/frommarek/83-atom-tau4/8-Ang/from_154K/6430/dos

  \end{figure}

Pseudogaps in the electronic density of states (EDOS) at the Fermi energy are confirmed based on our first-principle calculations in Fig.~\ref{fig:eDOS}, which agree with results reported in Ref.\cite{PHILLIPS1993}.

\begin{table}[htpb]
  \centering
  \begin{ruledtabular}
    \begin{tabular}{llllllll}
      index&$N_A$&$x_{\rm Al}$&$x_{\rm Co}$&$x_{\rm Cu}$&$\Delta E$ &Competing phases\\
      \hline                   
      Q0   &81&  0.593 &  0.148 & 0.259 &9.4&S4,S7,S11\\
      \hline
      Q1   &82&  0.646 &  0.171 & 0.183 &5.6&S4,S8,S11\\
      Q2   &82&  0.634 &  0.159 & 0.207 &6.0&S4,S8,S11\\
      Q3   &82&  0.646 &  0.159 & 0.195 &6.1&S4,S8,S11\\
      Q4   &82&  0.634 &  0.171 & 0.195 &6.8&S4,S8,S11\\
      Q6   &82&  0.646 &  0.146 & 0.207 &9.9&S4,S7,S11\\
      Q7   &82&  0.659 &  0.146 & 0.195 &10.9&S4,S7,S11\\
      \hline                     
      Q8   &83&  0.639 &  0.169 & 0.193 &3.0&S4,S8,S11\\
      Q9   &83&  0.627 &  0.193 & 0.181 &3.0&S4,S5,S9\\
      Q10  &83&  0.627 &  0.181 & 0.193 &3.3&S4,S8,S9\\
      Q11  &83&  0.639 &  0.181 & 0.181 &3.5&S4,S8,S9\\
      Q12  &83&  0.615 &  0.181 & 0.205 &4.3&S4,S5,S9\\
      Q13  &83&  0.627 &  0.169 & 0.205 &5.0&S4,S8,S11\\
      Q14  &83&  0.639 &  0.157 & 0.205 &5.4&S4,S8,S11\\
      Q15  &83&  0.651 &  0.181 & 0.169 &5.8&S4,S8,S11\\
      Q16  &83&  0.663 &  0.169 & 0.169 &5.2&S4,S8,S11\\
      Q17  &83&  0.614 &  0.193 & 0.193 &6.1&S4,S5,S9\\
      Q18  &83&  0.651 &  0.169 & 0.181 &6.6&S4,S8,S11\\
      Q19  &83&  0.651 &  0.157 & 0.193 &8.4&S4,S7,S11\\
      Q20  &83&  0.663 &  0.181 & 0.157 &10.6&S4,S8,S11\\
%      Q21&83&0.1566&0.2169&58.6&S4,S8,S11\\
      \hline
      Q21  &84&  0.6428      &0.1667&0.1905&5.8&S4,S8,S11\\
      Q22  &84&  0.6309      &0.1667&0.2024&9.7&S4,S8,S11\\
    \end{tabular}
  \end{ruledtabular}
  \caption{\label{tab:Hull-closeup} Compositions and $\Delta E$ (in meV/atom) of $a_3$ approximants above the convex hull}
  \end{table}

\begin{table}[htpb]
  \centering
  \begin{ruledtabular}
    \begin{tabular}{lll|lll}
      index&Approximants&$\Delta H$&index&Approximants&$\Delta H$\\
      \hline
      L0&Al$_{138}$Co$_{39}$Cu$_{40}$&11.8&L10&Al$_{138}$Co$_{40}$Cu$_{40}$&12.1\\
      L1&Al$_{139}$Co$_{39}$Cu$_{39}$&11.2&L11&Al$_{138}$Co$_{41}$Cu$_{39}$&11.2\\
      L2&Al$_{138}$Co$_{38}$Cu$_{41}$&10.0&L12&Al$_{138}$Co$_{39}$Cu$_{41}$&8.8\\
      L3&Al$_{137}$Co$_{39}$Cu$_{41}$&13.4&L13&Al$_{139}$Co$_{39}$Cu$_{40}$&10.9\\
      L4&Al$_{140}$Co$_{39}$Cu$_{38}$&17.8&L14&Al$_{137}$Co$_{39}$Cu$_{42}$&13.4\\
      L5&Al$_{141}$Co$_{38}$Cu$_{38}$&23.9&L15&Al$_{139}$Co$_{39}$Cu$_{40}$&10.9\\
      L6&Al$_{138}$Co$_{39}$Cu$_{41}$&8.8&L16&Al$_{138}$Co$_{40}$Cu$_{40}$&12.1\\
      L7&Al$_{138}$Co$_{41}$Cu$_{39}$&11.2&L17&Al$_{138}$Co$_{38}$Cu$_{42}$&13.0\\
      L8&Al$_{139}$Co$_{38}$Cu$_{41}$&8.5&L18&Al$_{139}$Co$_{39}$Cu$_{41}$&10.3\\
      L9&Al$_{137}$Co$_{40}$Cu$_{41}$&11.7&L19&Al$_{360}$Co$_{106}$Cu$_{103}$&11.4\\
    \end{tabular}
  \end{ruledtabular}
  \caption{\label{tab:1}Relative enthalpies $\Delta H$ [meV] of larger $a_4$ and $a_5$ approximants.}
\end{table}

\section{Approximant size}
\label{app:size}
\setcounter{figure}{0}

\begin{figure}[htpb]
  \centering
%  \begin{subfigure}[b]{.45\textwidth}
    \centering
    \includegraphics[width=.45\textwidth]{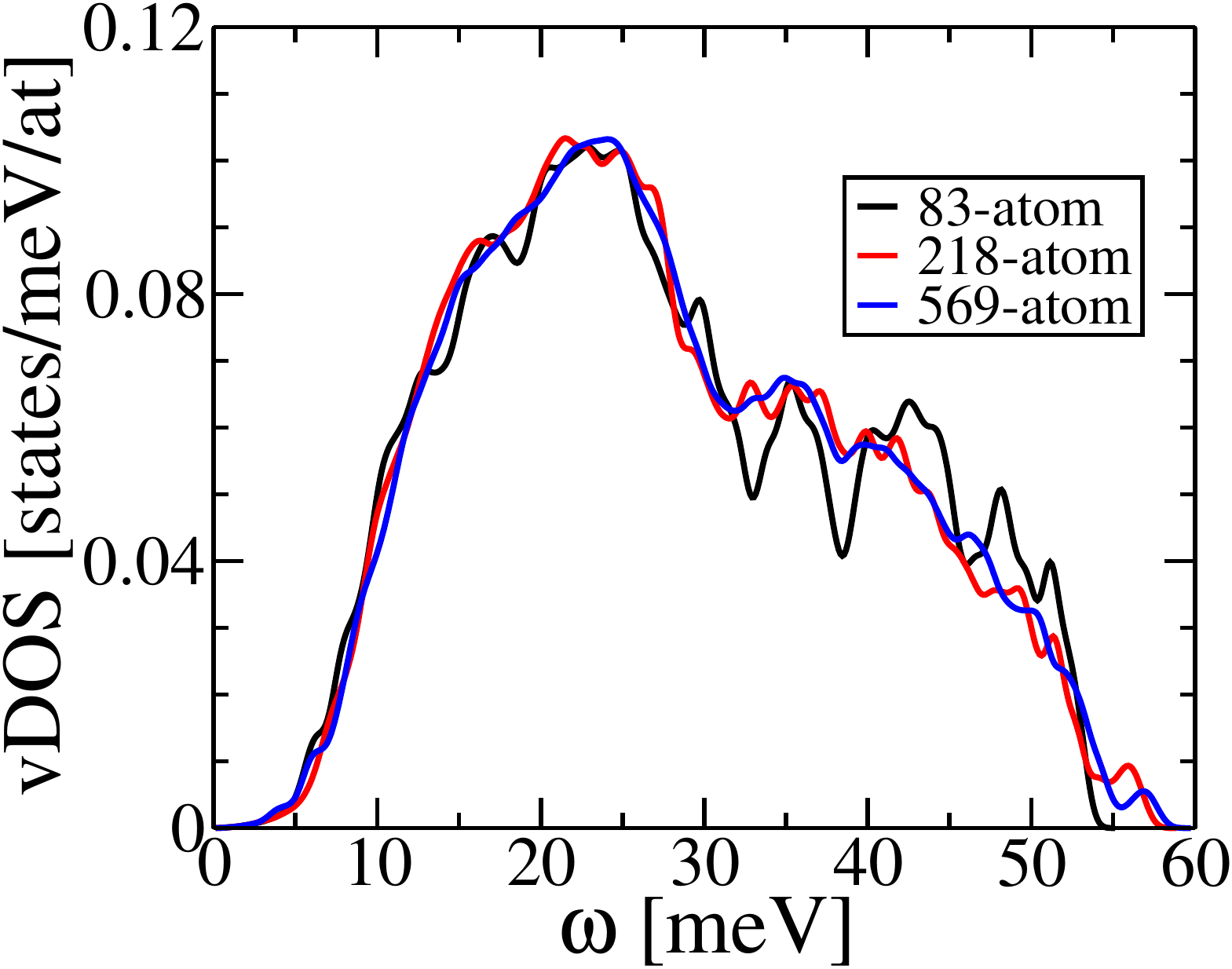}
%    \caption{\label{fig:13a}}
%  \end{subfigure}
  % \begin{subfigure}[b]{.2\textwidth}
  %   \centering
  %   \includegraphics[width=\textwidth]{fvib_vs_N_qc.pdf}
  %   \caption{\label{fig:13b}}
  % \end{subfigure}
  \caption{  \label{fig:vDOS_size} Vibrational density of states of Q8, L0 and L19 calculated using EOPP}
\end{figure}

\begin{figure}[htpb]
  \centering
  \includegraphics[width=.23\textwidth]{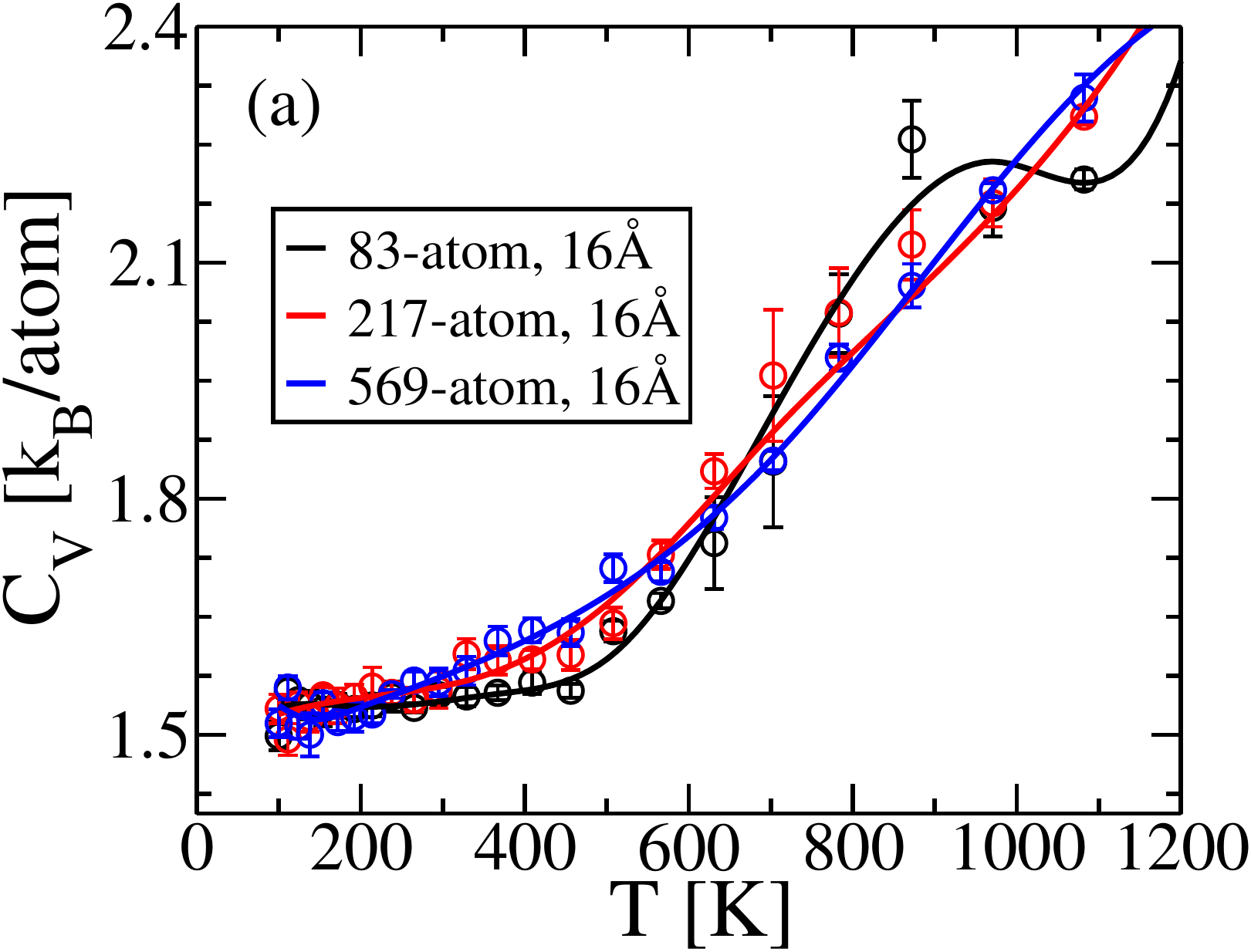}
  \includegraphics[width=.23\textwidth]{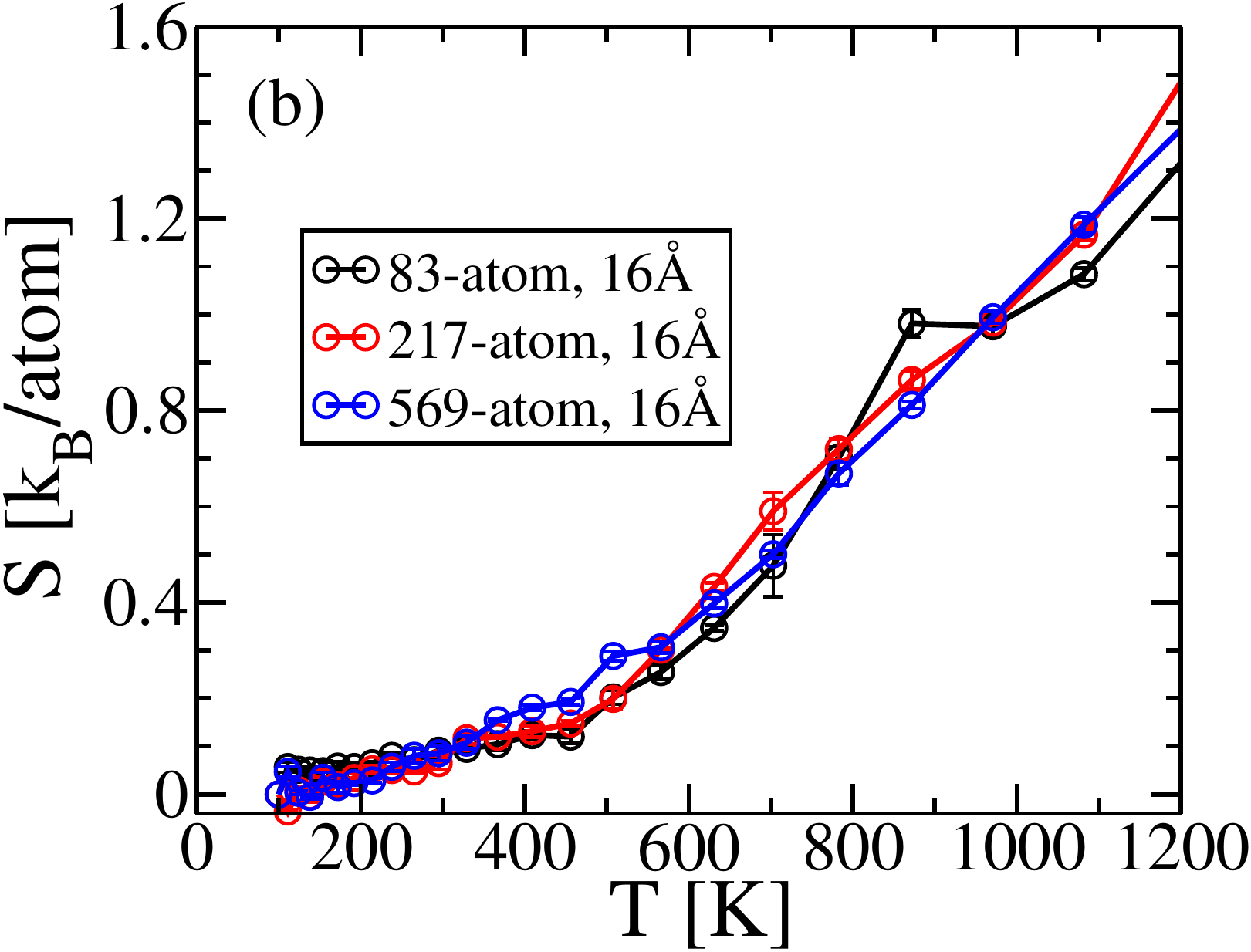}
  \includegraphics[width=.23\textwidth]{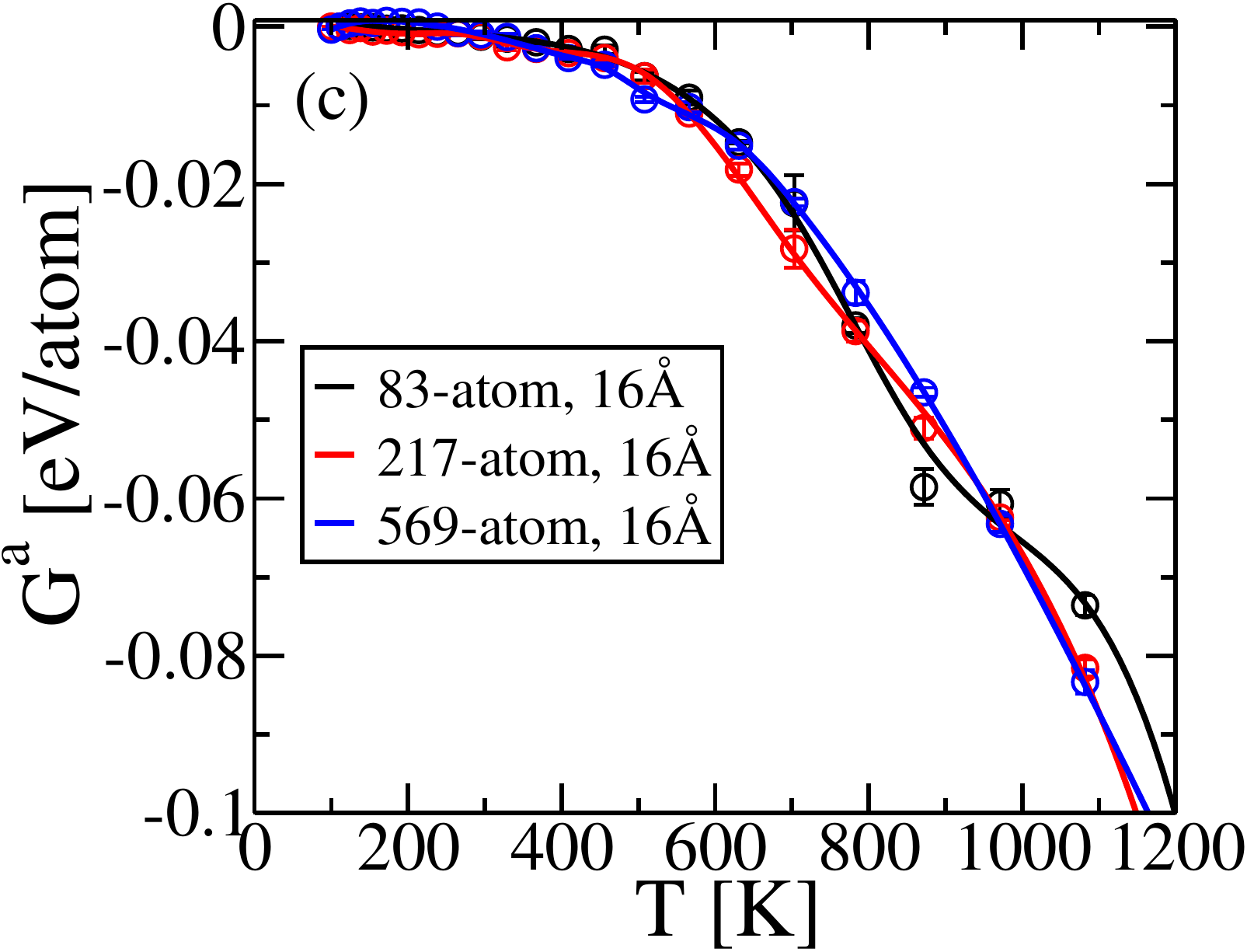}
  \includegraphics[width=.23\textwidth]{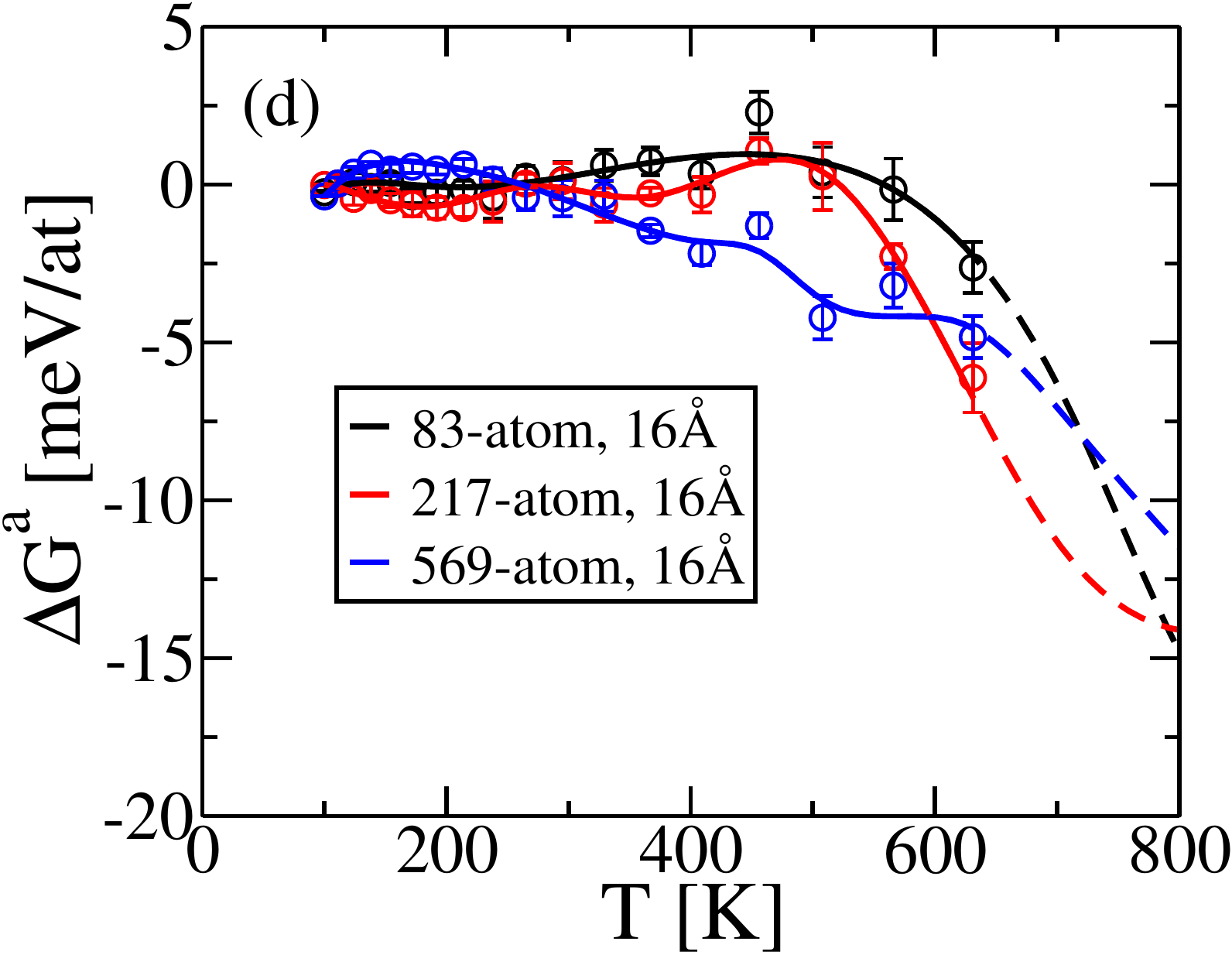}
  \caption{  \label{fig:thermo_size}
    (a) Heat capacities $c_v$, (b) entropies $S$, (c) anharmonic free
    energies $G^a$ and (d)
    relative anharmonic free energies $\Delta G^a$ of the 83-atom, 
    217-atom and  569-atom QCAs.}
  \end{figure}

\begin{figure}[htpb]
  \centering
  \includegraphics[width=.3\textwidth]{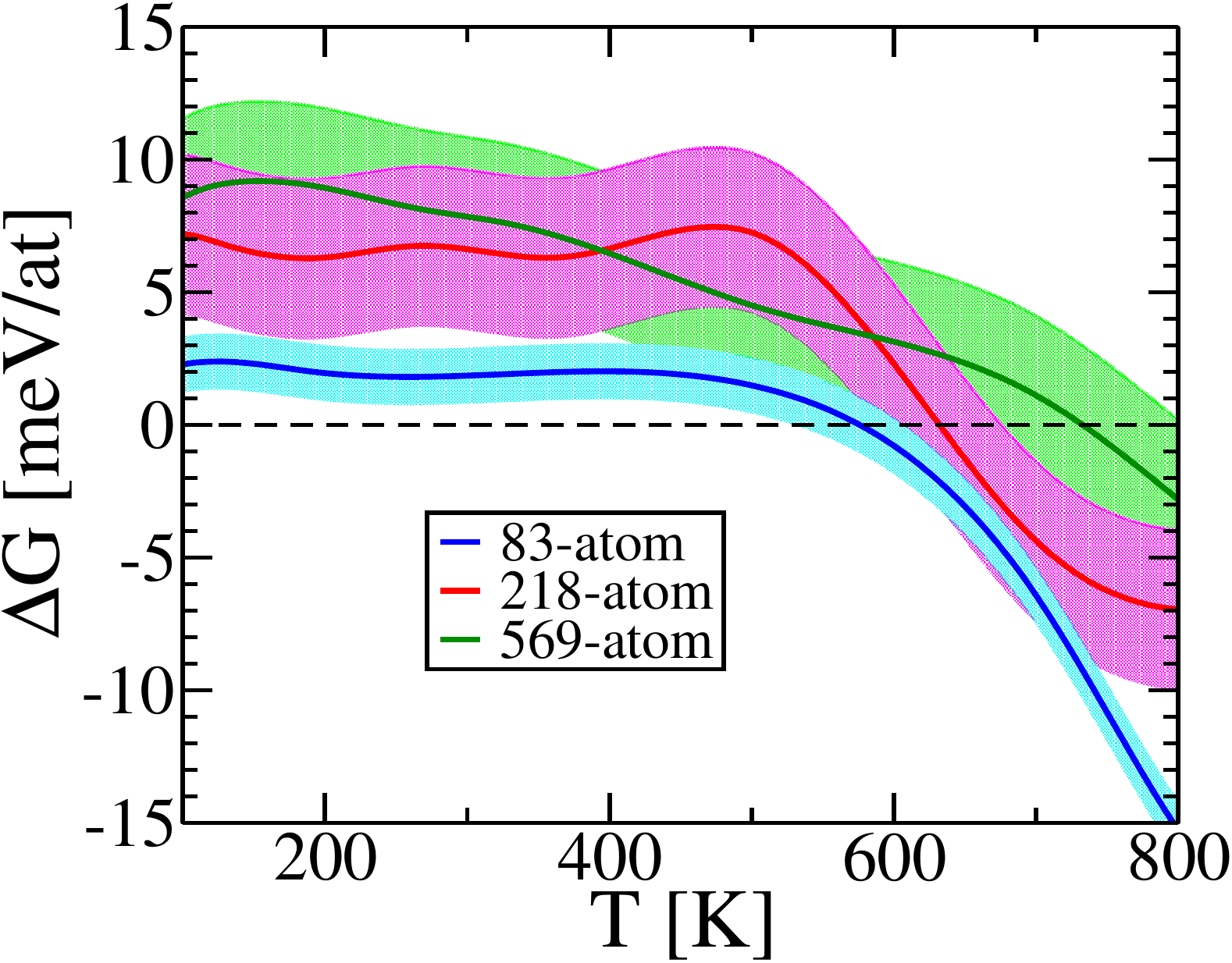}
  \caption{  \label{fig:dG_size}
    Total free energies $\Delta G$ and predicted transition
    temperature $T$ for three sizes. Uncertainties are indicated by
    pink and green strokes with $\sigma_G\sim 3$meV according to
    differences in vibrational free energies (see Appendix~\ref{app:EOPP}).}
\end{figure}

Many properties vary depending on approximant size. Happily, the vDOS as calculated within EOPP are very close (see Fig.~\ref{fig:vDOS_size}), and the harmonic free energies are equal to within 1 meV/atom over the temperature range of interest. Fig.~\ref{fig:thermo_size} shows the size dependences of other thermodynamic properties, and Fig.~\ref{fig:dG_size} shows the cumulative $\Delta G$, including all contributions for various QCA sizes. Owing largely to our inability to identify large low energy approximants, the stabilization temperature for the QCA relative to its competing crystal phases grows from 600 to 800K.

All of our free energy calculations are carried out within structures with 16~\AA~ periodicity. Entropy at 8A was 10 meV/atom lower for all approximant sizes.

\section{More about $\tau$ (hP5) and B2 phases}
\label{app:tau}
\setcounter{figure}{0}

\begin{figure*}[htpb]
  \centering
  \includegraphics[width=0.3\textwidth]{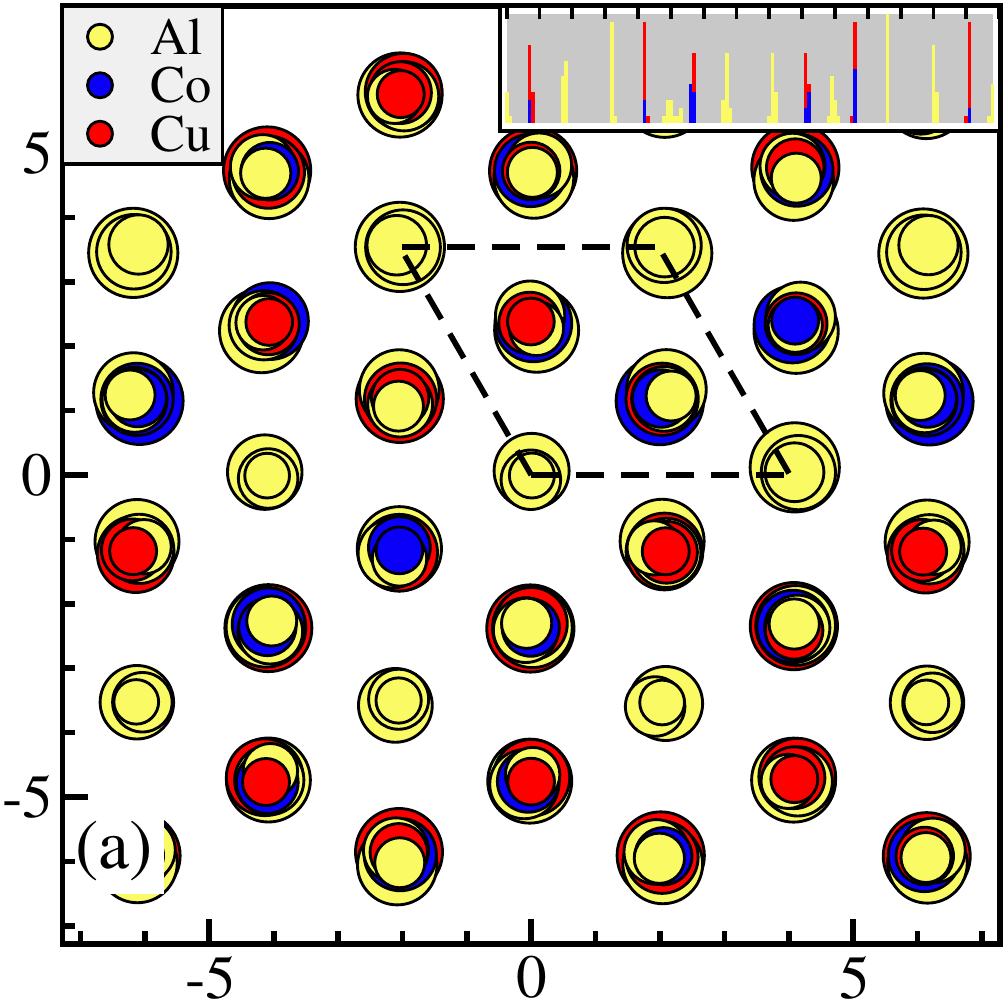}
  \includegraphics[width=0.3\textwidth]{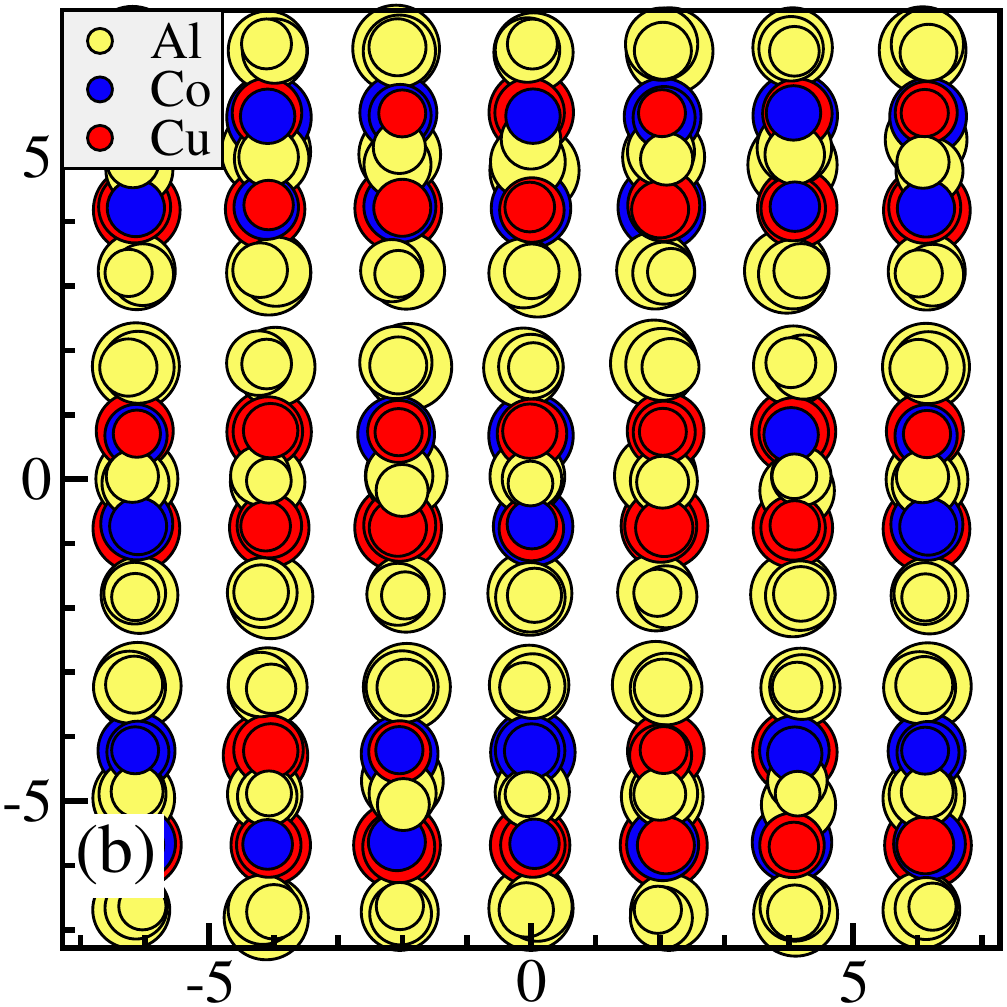}
  \includegraphics[width=0.3\textwidth]{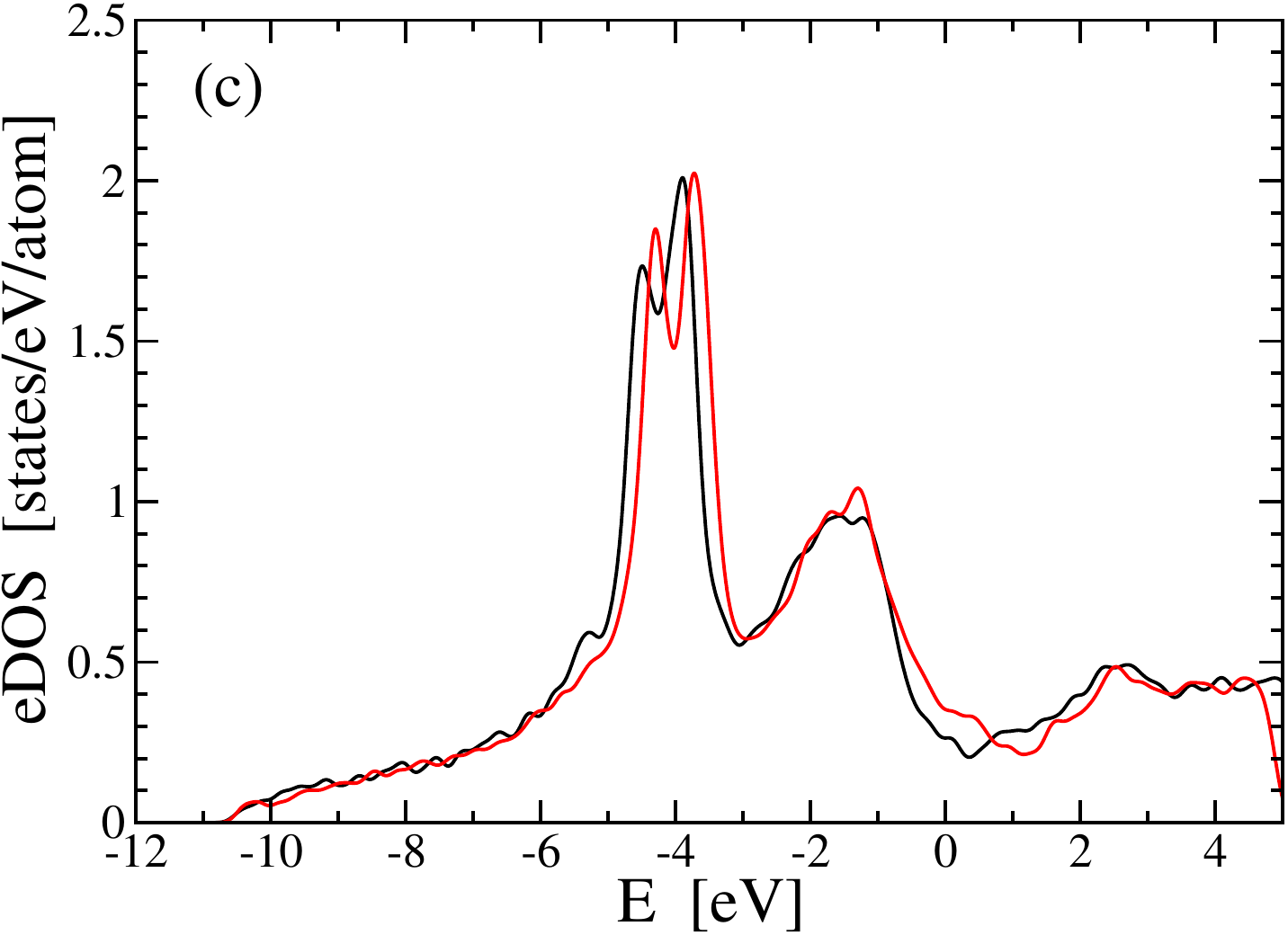}
  \caption{(a) six-fold axis and (b) side view of a VASP-relaxed random 3x3x3 supercell of $\tau_3$ (hP5) Al$_{6}$Co$_{1.5}$Cu$_{2.5}$. Unit cells are denoted by dash black lines. (c) eDOS smeared by 0.05 eV.}
  \label{fig:hP5}
\end{figure*}

The $\tau$-Al$_3$(Co,Cu)$_2$.hP5 family~\cite{widom2000,vanSande1978,VANTENDELOO1989} is a vacancy-ordered phase derived from B2. The vacancies assemble periodically in the transition metal (TM) (111) planes. The phase $\tau_n$ has periodicity $n$, where $n$ is a Fibonacci number.  Fig~\ref{fig:hP5} illustrates the $\tau_3$ phase, Al$_{12}$Co$_{3}$Cu$_{5}$, in which every third TM layer is vacant. When subjected to our replica exchange MCMD simulation, this structure transformed to B2 with vacancies. That is, some of the TM atoms moved to the vacant layer. This transformation occurs more quickly at high temperatures, where TM atoms diffuse through the Al layers. In the process, the EOPP energy dropped by 22 meV/atom, while the relaxed VASP energy rose by 57 meV/atom.
% Yang: please check these numbers. {\bf (xx: These are right. See /home/hyhy123/alloy.pbe/AlCoCu/Al6CoCu3.hP5/snapshot/splits/widom/vacant-cP2-from-temper/run0, /home/hyhy123/alloy.pbe/AlCoCu/Al6CoCu3.hP5/snapshot/splits/widom/my-hP5)}
Noting the significant reduction in the Fermi level density of states in the vacancy-ordered structure relative to B2 with vacancies, we tentatively attribute this disagreement to an electronic structure effect that was not captured by our EOPP potentials.

Owing to the instability of $\tau_3$ under our simulation, we are unable to calculate its anharmonic free energy in the manner of the other phases. Instead, we perform short runs at low temperatures, where the phase remains metastable, then we extrapolate to estimate the free energy at high temperatures. The exptrapolated data are shown as dashed lines in Fig.~\ref{fig:5}.

\section{Phason flips}
\label{app:phason}
\setcounter{figure}{0}

The peak in heat capacity of the 1x1x2 (16~\AA) $a_3$ QCA seen in Fig.~\ref{fig:thermo_size}a is due to unlocking of interlayer phason fluctuations. Below the peak, the structure has well-formed PB motifs with five-fold symmetry and 8~\AA~ stacking periodicity. Flat and puckered layers alternate; the puckered layers are mirrored across the flat layers. Notice that the flat layers are relatively poor in Cu, as exemplified by the PB motifs in Fig.~\ref{fig:mC102}b and~\ref{fig:83}c. Additionally, the structure has 5-fold symmetry, because the pentagon orientations of the PB are preserved through the entire vertical stack. The positions of the PB within a plane, and the vertical height of the PB equators, is preserved throughout a long MCMD run.

\begin{figure*}[htpb]
  \centering
  \includegraphics[width=0.45\textwidth]{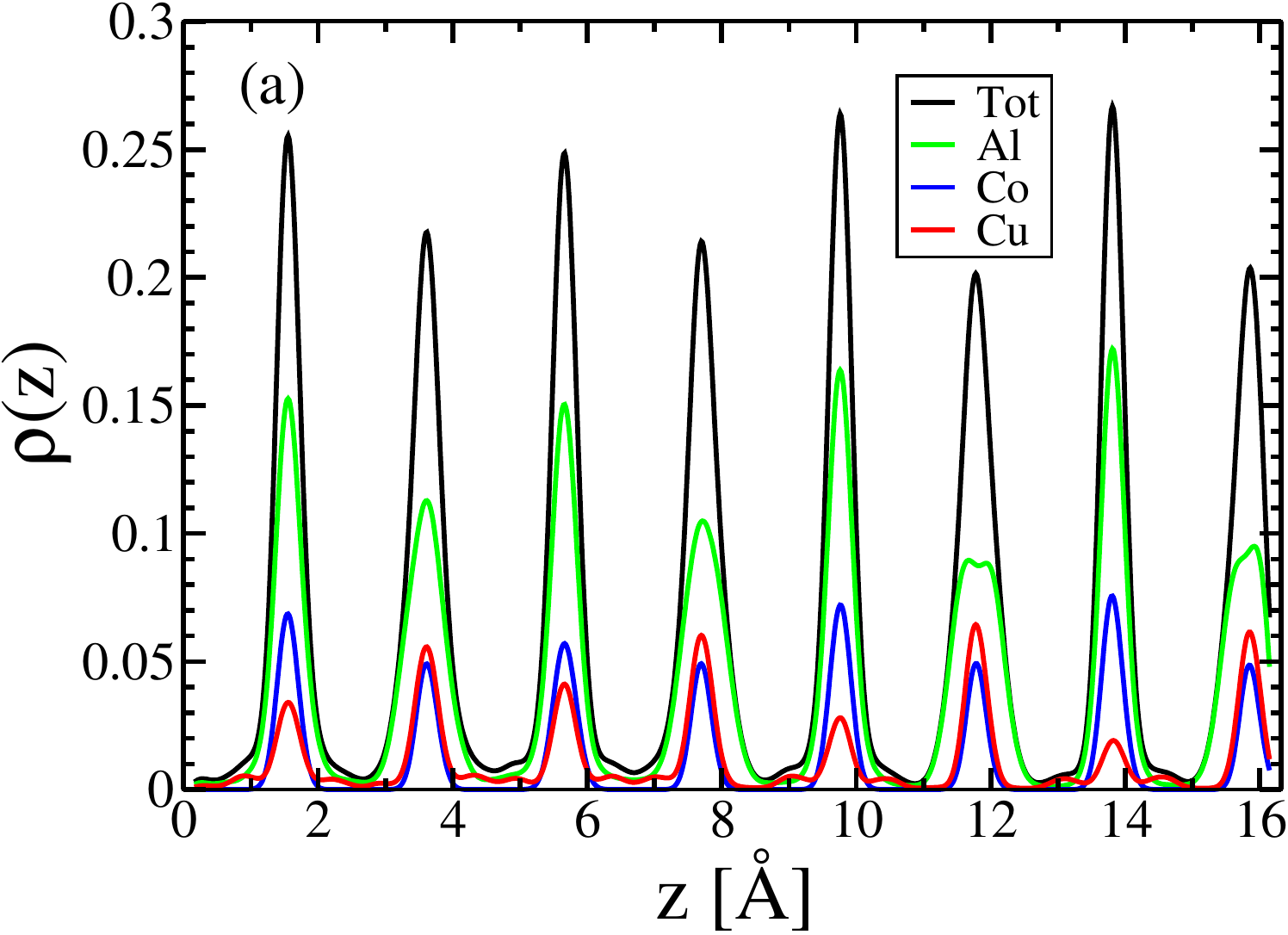}
  \includegraphics[width=0.45\textwidth]{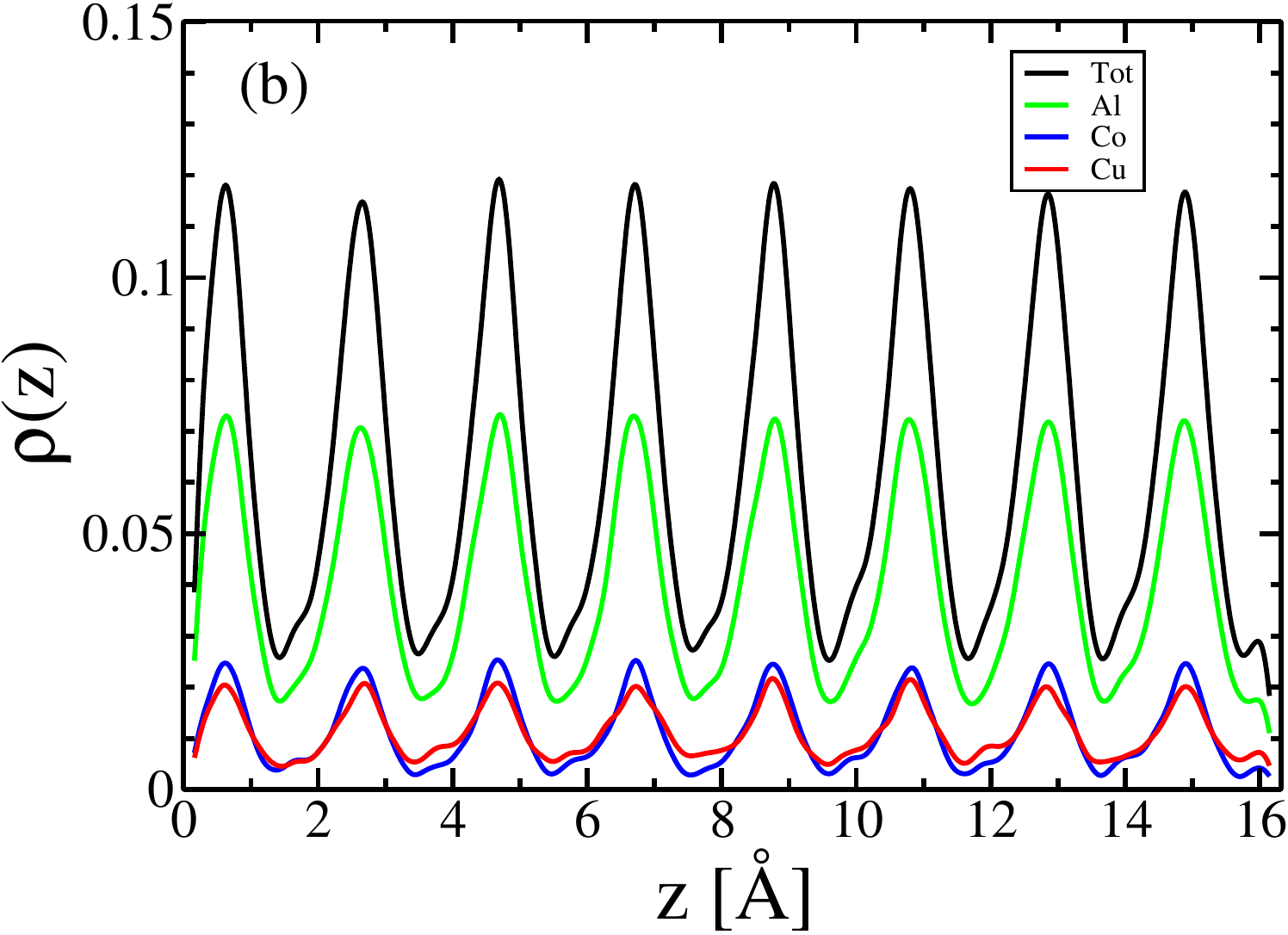}
  \caption{  \label{fig:8to4}
    $\rho(z)$ distributions (a) below (T=703K) and (b) above (T=971K) the heat capacity peak of our 1x1x2 supercell of $a_3$ QCA.}
\end{figure*}

Above the peak, the distinction between flat and puckered is lost on average. Both flat and puckered motifs coexist in a given layer. Motifs in adjacent layers are rotated by 36$^\circ$, creating a $10_5$ screw axis. Throughout the MCMD run, the positions of PB equators, junction layers and caps migrate both within each layer and between layers. Complete PBs form occasionally, even spanning the full 16~\AA~ stack, but usually only PB fragments are observed.

\begin{figure*}[htpb]
  \centering
  \includegraphics[width=0.3\textwidth]{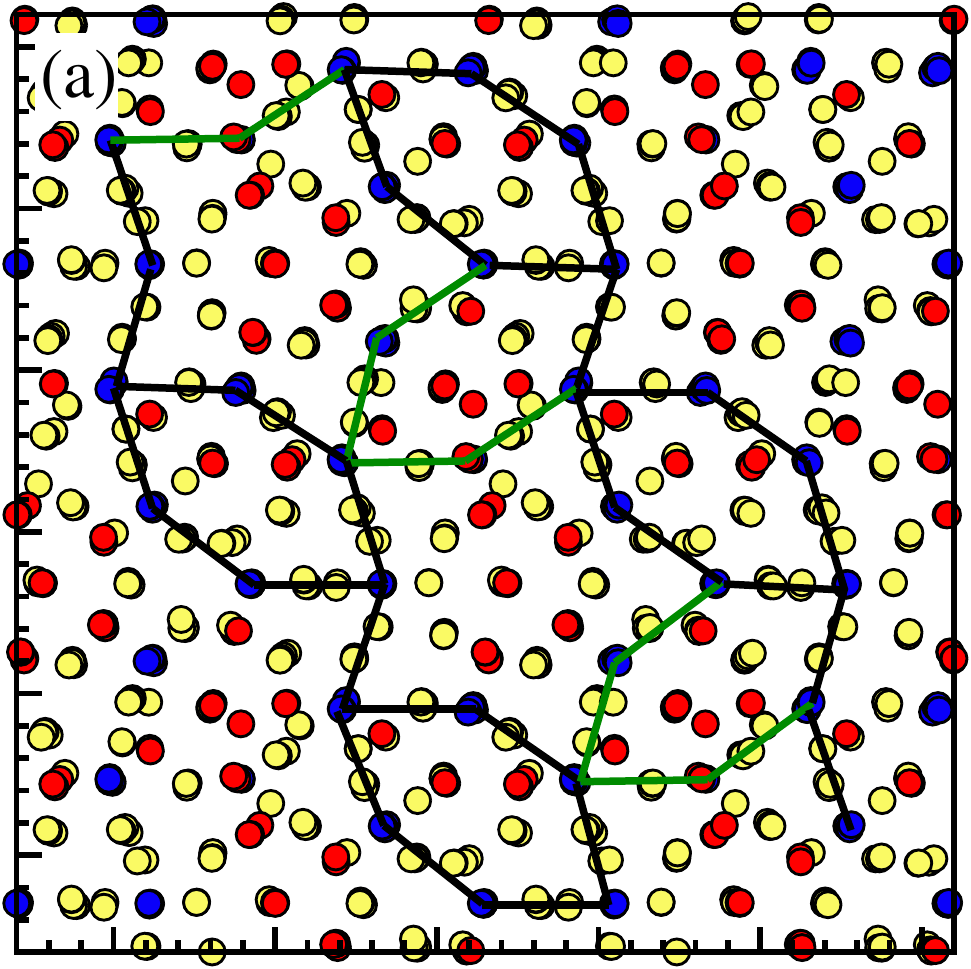}
  \includegraphics[width=0.3\textwidth]{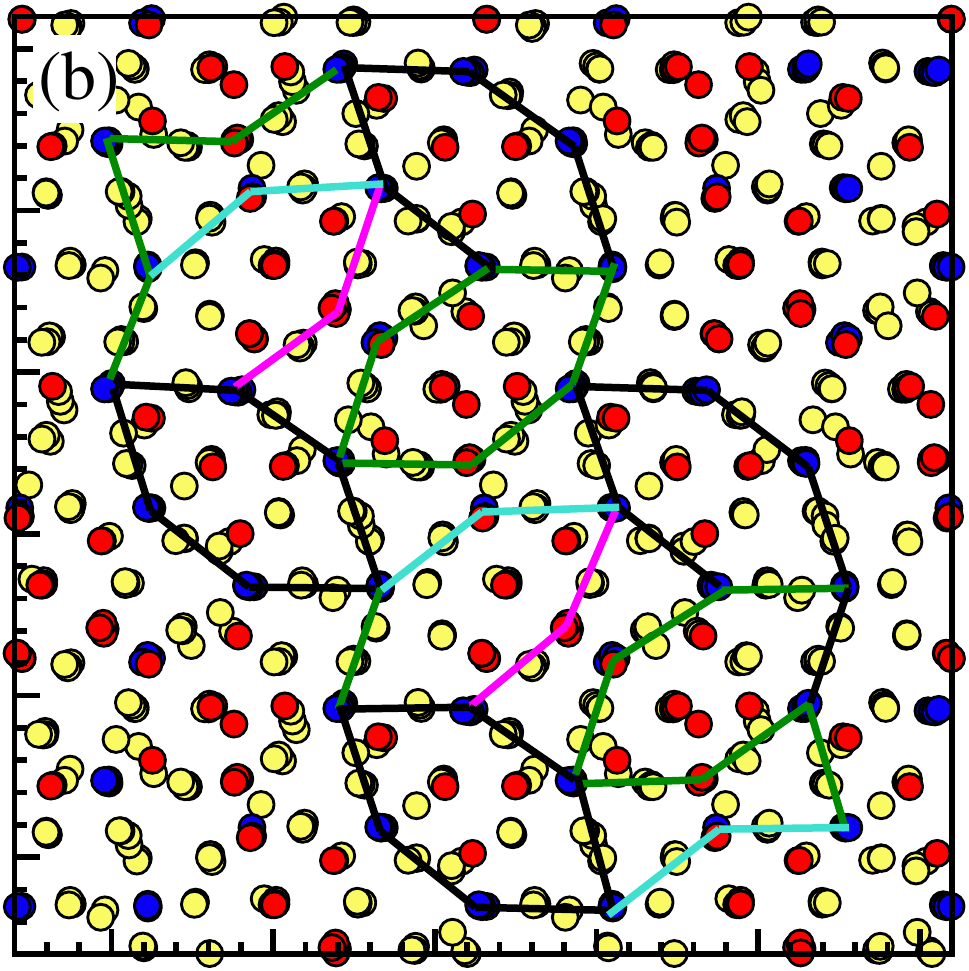}
  \includegraphics[width=0.3\textwidth]{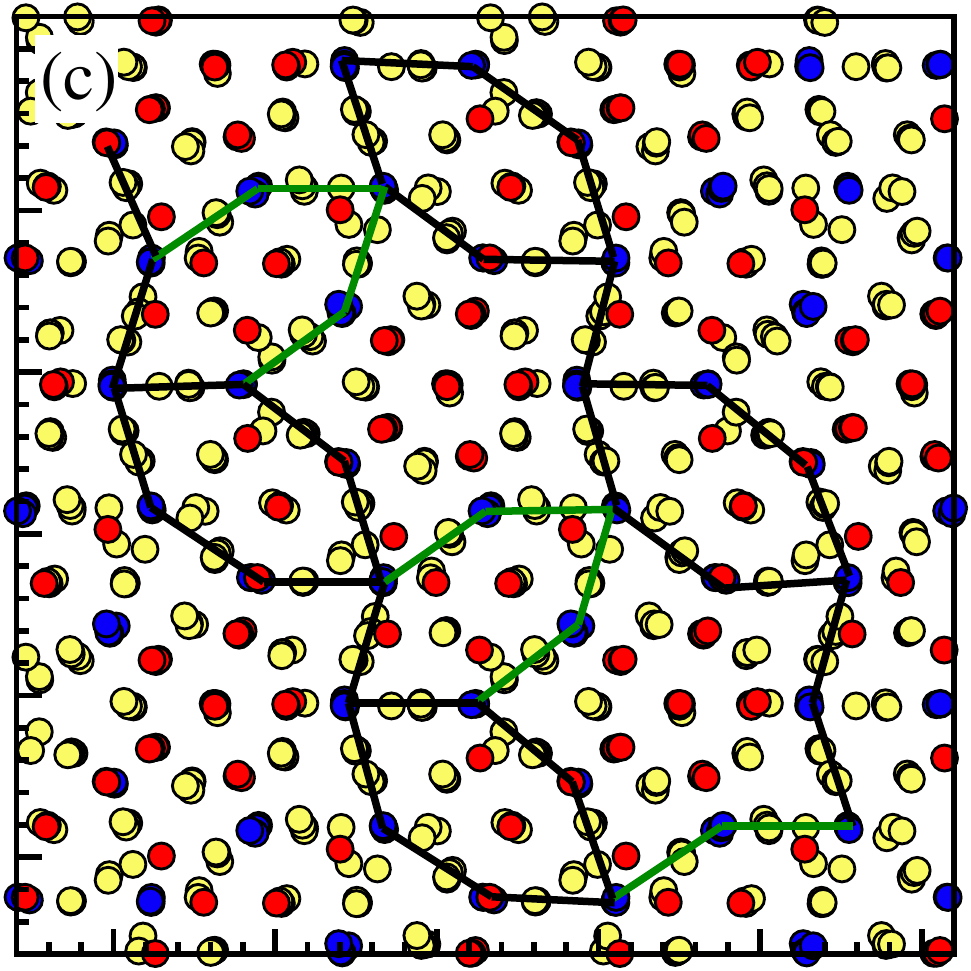}
  \caption{Tiling sequence for phason flips in 1x1x2 (16~\AA) supercell of $a_3$ approximant at T=971K. Structures have been relaxed under EOPP. (a) and (c) are actual tilings produced by connecting Co atoms centering PB caps. (b) illustrated phason flip sequence.}
  \label{fig:phason}
\end{figure*}

To illustrate the evolution of structure, Fig~\ref{fig:phason} shows three structures separated by short intervals of MCMD at 971K. The structures have been relaxed under EOPP and projected into a single plane. In parts (a) and (c), Co-Co bonds of 3.9-4.1~\AA are drawn; in general these form HBS tilings. Bonds in black are shared in the two configurations, while bonds in green differ. Note that each tiling contains a star and two hexagons, but they differ in the positions of these tiles and in the orientation of the stars. Part (b) shows a sequence of ``bowtie'' flips~\cite{Bowtie2006} that transform (a) into (c). First, the pair of bonds in magenta form, replacing the adjacent green bonds. This converts the H2S tiling into a B2 tiling consisting of a pair of boats. Next, the turquoise bonds replace the adjacent green bonds, resulting in the new H2S tiling of part (c). Since the new star has reversed its orientation, such transformations convert five-fold symmetric structures into ten-fold. A video illustrating continuous time evolution is presented in the Supplemental Material.

\begin{figure}[htpb]
  \centering
  \includegraphics[width=.45\textwidth]{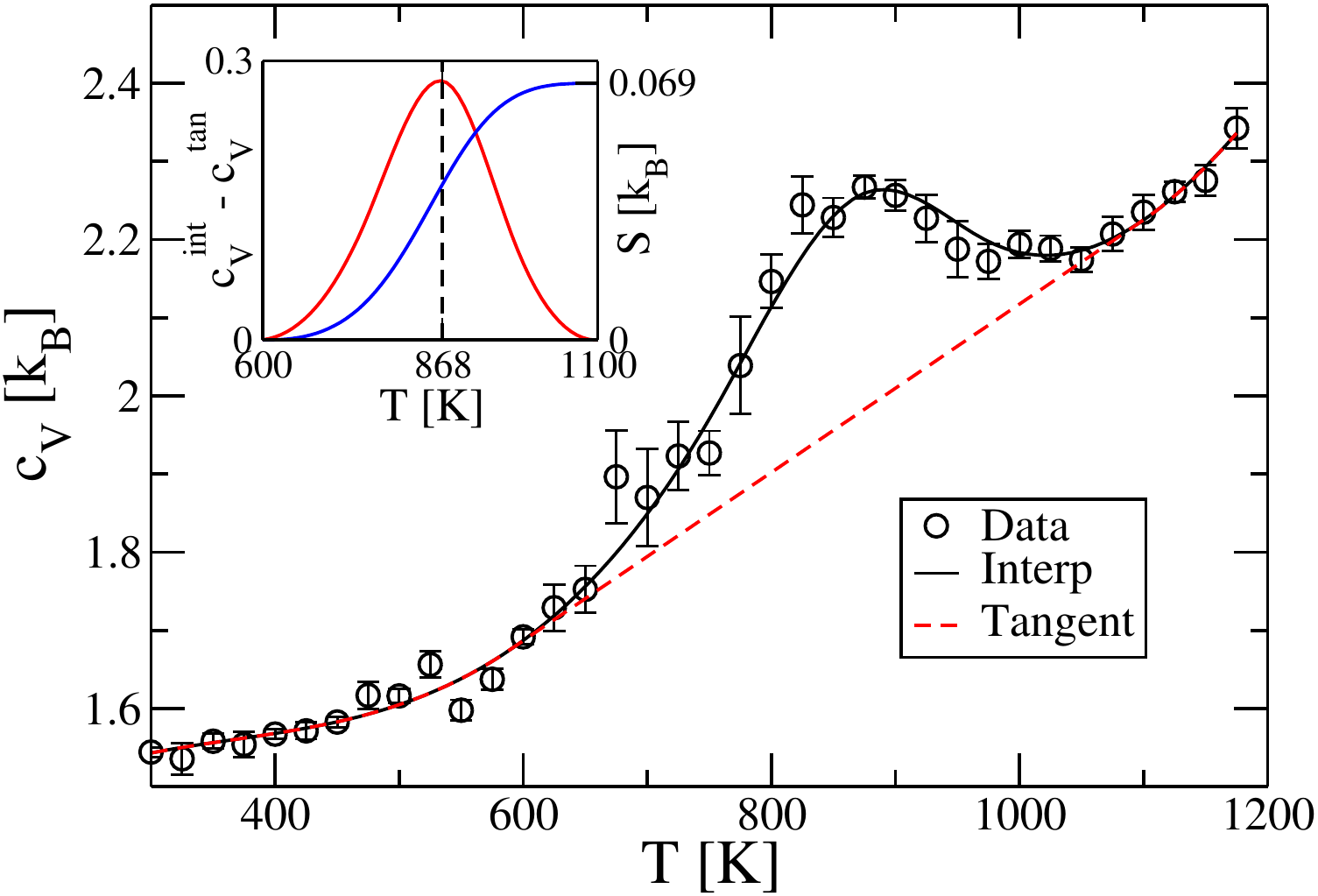}
  \caption{Specific heat capacity $c_V$ of 83-atom $a_3$ approximant. Inset shows excess heat capacity $\Delta c_V$ relative to tangent line and the integrated entropy.}
  \label{fig:sphason}
\end{figure}

In practice each bowtie flip consists of a sequence of individual steps in which each bowtie in each layer transforms individually. In fact, each is associated with a sigle Co/Cu swap. In our H2S tiling there are five possible bowtie flips per flat layer, two of which can occur (nearly) independently. In each case there are three possible states (unflipped, swapping with puckered layer above, and swapping with layer below), resulting in a net of 30 states per flat layer, or $30^2$ per 83 atoms. We estimate an entropy of
\begin{equation}
  S_{\rm Phason} \approx \frac{1}{83}\kB\ln{30^2} = 0.082\kB
\end{equation}
that works out to $TS_{\rm Phason}\approx 7$ meV/atom at $T=1000K$. This estimate is approximately confirmed by thermodynamic integration of the heat capacity peak seen in Fig.~\ref{fig:thermo_size}a and enlarged in Fig.~\ref{fig:sphason}. We evaluate the entropy gain due to phason unlocking by integrating the excess heat capacity, $\Delta c_V$,
\begin{equation}
  \label{eq:DS}
  \Delta S = \int_{600}^{1100}\rmd T \frac{c_V}{T} = 0.069\kB.
\end{equation}

\end{appendix}

\bibliography{refs}
\end{document}